\documentclass[a4paper, 11pt]{article}
\usepackage{amsmath, amsfonts, amsthm, mathrsfs, graphicx, cite}

\title{The Mathematical Syntax of Architectures\footnote{The very first version of this paper has been called \emph{The Logical Syntax of IT Architectures} with an intended allusion to Richard Carnap's seminal work \cite{Carnap-1934}. However, during our further studies we have abandoned the characterization of our approach as \emph{logical} because the vast amount of existing research around \emph{Architecture Description Languages} (ADLs) and their accompanying (typically first order) logics might be more appropriately called \emph{logical syntax} of architectures. We now call our foundational work a \emph{mathematical} syntax because it purely relies on mathematical concepts. We also dropped the ``IT'' classifier to highlight the universal applicability of our formalization (including, for instance, systems engineering).}
}
 
\author{Christoph F. Strnadl\footnote{contact: Gaia-X CTO team, \texttt{christoph.strnadl@gaia-x.eu}; Twitter: \texttt{@archimate}}\;, IEEE \emph{Senior Member}}

\date{November 30, 2024}

\newcommand{\mA}{\mathscr{A}}
\newcommand{\mB}{\mathscr{B}}
\newcommand{\mC}{\mathscr{C}}
\newcommand{\mI}{\mathscr{I}}
\newcommand{\mM}{\mathscr{M}}
\newcommand{\mT}{\mathscr{T}}

\begin{document}

\maketitle

\begin{abstract}
Despite several (accepted) standards, core notions typically employed in information technology or system engineering architectures lack the precise and exact foundations encountered in logic, algebra, and other branches of mathematics. 

In this contribution we define the syntactical aspects of the term \emph{architecture} in a mathematically rigorous way. We motivate our particular choice by demonstrating (i) how commonly understood and expected properties of an architecture---as defined by various standards---can be suitably identified or derived within our formalization, (ii) how our concept is fully compatible with real life (business) architectures, and (iii) how our definition complements recent foundational work in this area \cite{Wilkinson-2018, Dickersen-2020, Efatmaneshnik-2020}.

We furthermore develop a rigorous notion of architectural \emph{similarity} based on the notion of \emph{homomorphisms} allowing the class of architectures to be regarded as a category, \textbf{Arch}. We demonstrate the applicability of our concepts to theory by deriving theorems on the classification of certain types of architectures. \emph{Inter alia}, we derive a \emph{no go} theorem proving that, in contrast to $n$-tier architectures, one cannot sensibly define generic \emph{architectural modularity} on the syntactical level alone.  
\end{abstract}

\tableofcontents

\section{Introduction}


Information technology architecture or systems architecture is a fun\-da\-men\-tal and widely deployed construct used when designing software- or tech\-no\-lo\-gy-intensive systems. Several architectural standards and frameworks are known and widely accepted: ISO/IEC/IEEE~\cite{ISO-42010}, TOGAF~\cite{TOGAF}, OMG UML~\cite{UML} and SysML~\cite{SysML}, ArchiMate~\cite{Archimate, Lankhorst-2017}, and others.

Nevertheless, the formalization	 attempts of the core entities of all frameworks or standards---as of to-date---remain on an essentially non-mathe\-matical and non-exact level using ordinary (i.e., every day and prose) language.\footnote{OMG's OCL (object constraint language) \cite{OCL} comes close to a rigorous language but is still limited to operating on the otherwise undefined core notions of UML.} 

Interestingly, this is in stark contrast to many other disciplines where similar endeavours of formalization were undertaken over 100 years ago, e.g., mathematics~\cite{Whitehead-Russell-1910}, philosophy~\cite{Wittgenstein-21}, language~\cite{Carnap-1934, Chomsky-1975}, architecture \emph{proper}~\cite{Mitchell-1990}, or even poetry~\cite{Hamburger-1987}.

This situation has recently also been recognized by the International Standards Organisation (ISO) whose JTC1/SC7/WG42 working group on System Architecture began a review of the currently adopted (prose) definitions in 2018. However, no (new) standard has yet been issued.

In order to contribute to the resolution of this unsatisfactory \emph{status quo},
this paper introduces a rigorous framework capturing the essence of the mathematical syntax of architectures. Note that we deliberately focus on the syntactical side of architecture omitting any discussion and treatment of the actual semantics of an architecture.

After a recapitulation of previous related work (section \ref{sec-RelatedWork}) we proceed in a highly deductive way and firstly present (in section \ref{sec-Theory}) the theoretical underpinnings of our theory. The applicability of our theory is then considerably expanded in section \ref{sec-Application} where we show that our approach is fully compatible with and able to serve as the very foundation of classical architecture standards. We also apply our formalism to current foundational work \cite{Dickersen-2020, Wilkinson-2018,Efatmaneshnik-2020} and succeed in providing a rigorous mathematical formalization of concepts employed therein. The section is complemented by several (mathematical) theorems on the classification of architectures demonstrating that our purely syntactical theory has a merit of its own even without any reference to architectural semantics. \emph{Inter alia}, we derive an interesting "no go" theorem that one cannot sensibly define the concept of \emph{modularzation} (i.e., the segmentation of a system into a disjunct set of constituting modules) on the level of syntax alone. This is in stark contrast to the case of $n$-tier architectures where this, as we are able to show, is possible. A concise discussion in section \ref{sec-Discussion} provides additional links to closely related topics and points to potential further (future) work using our approach.

\section{Related Work}
\label{sec-RelatedWork}

Considerable research has been expended in identifying proper ways of formalizing (software) architecture with a lot of effort going into the definition of suitable \emph{Architecture Description Languages} (ADLs). In addition to providing a (typically first-order logic based) concrete and strict syntax for modelling architectures, they typically add further functionality pertaining to displaying, analyzing, or simulating architectural descriptions written in a specific ADL \cite{Garlan-2003}.

Even before that, Category Theory has been put forward by some as a rigorous but enormously flexible methodology for computing science \cite{Goguen-1989}. This branch of mathematics has sometimes been used to supplement classical ADL-based approaches \cite{Fiadeiro-1996, Wermelinger-1999}. 

However, Category Theory and architectural practice do not seem to match perfectly in many circumstances 
as the hurdle between concrete architectural work (e.g., conceiving, describing, and communication an architecture) and the intentionally highly abstract level of Category Theory remains large\footnote{For instance, in order to formalize the---quite simple---idea that some components consist of other (sub-)components one needs (co-)limits right at the very heart of the theory.} Still, one very attractive concept here is the generalized notion of \emph{morphisms} (including \emph{isomorphisms}), that is structure preserving maps, which we will re-use in an appropriately specialized form.

Within service-oriented architectures (SOA) \cite{Strnadl-2006} the vagueness of the term \emph{service} (which is almost universally defined in prose only) has prompted some authors to introduce a more rigorous formalization of their SOAs. Asha \emph{et al.{}} \cite{Asha-2020} define a (web) service as a relation over the set of observable (service) properties and functionalities (which they call ``appearances''). However, this definition is not further utilized in their contribution. 

Yanchuk \emph{et al.{}} \cite{Yanchuk-2006} define a (logical) \emph{service instance} $S$ in terms of its functionality $f$, its interface $I$ and the set of ``coordinated and interfacing'' processes $P_1, P_2,...,P_k$ actually implementing $S$: $S = \langle \, P_1, P_2,...,P_k, \Lambda \, \rangle$. A process $P_i = P_i(f,I)$ implements the functionality $f$ of $S$ and exposes it through interface $I$. $\Lambda$ is the so-called ``network communication function'' between the individual processes. Recognizing SOA's fundamental composition principle, where finer grained services may be aggregated\footnote{This is often called ``orchestration'' in distinction to ``choreography''. The latter term refers to the coordination (in time) of a set of distinct services without a governing entity (such as an enterprise service bus or business process engine).} into larger services, they define an \emph{application} $A$ as a directed graph with edges $E$ over a set of \emph{service instances}, $A$ = $\langle \, V_A = \{S_1, S_2,...,S_k \}, E \subseteq V_A \times V_A \, \rangle$. This formalization is then used to derive and rigorously define several ``application classes'', e.g.~facade and satellite applications, transient applications, accumulating applications, and others.

A similar, but less refined set-theoretic approach is also found in \cite{Goikoetxea-2004}. There, an \emph{enterprise architecture} $\mathscr{E}$ is essentially defined as an otherwise unstructured 8-tuple $\mathscr{E} = \langle \, R, B, S, D, A, T, C, M \, \rangle$ comprising the set of requirements $R$, business processes $P$, business systems $S$, data elements $D$, applications $A$, technologies $T$, constraints, metadata, design rules $C$, and the set of architectural metrics $M$. Due to the complete lack of any further structure besides the categorization of architectural elements into 8 sorts, this definition does not convey any syntactical information.

The requirement of being able to transform architectural models into one another has been evident since the very beginnings of architecture work and lies at the heart of OMG's \emph{model driven development} (MDD) and the succession of models contained in this approach. A formalization of this transformation, though, is not provided within MDD itself. Attempts to do so recognize that, in order to be able to formally (later then also automatically) transform one model into another, one has to specify the vocabulary of the two models, and the syntax they possess \cite{Lopez-2009}. In this reference, the transformation of a source model into a target model is specified as a so-called ``weaving model'' capturing the ``correspondence'' between the different model elements. In our own contribution, we will rigidly formalize this notion.

From a decidedly logical point of view, the concept of a \emph{signature} of a (formal) language $L$ has been used to ``define'' the syntax of architectures \cite{Broy-1995, deBoer-2004}. In pure model theory \cite{Doets-1996, Hodges-1993}, a \emph{signature} 
$\sigma = \langle \, R, F, C \, \rangle$ defines the set of non-logical symbols of $L$, i.e., its vocabulary consisting of the set $R$ of relation symbols, $F$ of function symbols, and constants $C$. Translated to computer science, a \emph{signatures} specifies the \emph{name space} of an architecture; the set of constants $C$ is then equated with (or replaced by) the set of sorts $S$.

However, without any additional structure or further information, a signature alone does not contain any \emph{syntactical} information at all because $\sigma$ does not restrict
the way how \emph{terms} of a given language $L$ are formed. Or to put it otherwise: A (such understood) signature of an architecture does not prescribe any syntax at all.
Hence, the claim of having defined the ``syntax of an architecture'' by solely specifying $\sigma$ that is providing the names of the architecture elements and just the names of the relations they potentially have to each other fails. Recognizing this, de Boer \emph{et al.}~\cite{deBoer-2004} introduce a partial order on the set of (primitive) sorts $S$ and on the set of relation symbols $R$. Thereby they are able to capture certain ``ontological'' aspects of an architecture like generalization, aggregation, or containment.

Even though endowing architectures (e.g., diagrams or other representations) with precise meaning has always been a key concern of architecture research and practice alike \cite{Garlan-2003}, it is important to recall that this contribution focuses on the \emph{syntactical} parts as opposed to the \emph{semantic} aspects of an architecture. While this distinction between syntax and semantics is often recognized (e.g.,\cite{Broy-1995, Astesiano-2003, deBoer-2004, Dickersen-2020}), the focus very often lies on the semantic elements with only a rudimentary or narrow formalization of the underlying syntax. In this sense, our work is complementary to the semantic ``thread'' within the  architecture domain.

\section{Theory}
\label{sec-Theory}

\newtheorem{definition}{Definition}
\newtheorem{theorem}{Theorem}	
\newtheorem{lemma}{Lemma}
\newtheorem{corollary}{Corollary}

\subsection{Basic Definitions}

Let us start directly with the definition of an \emph{architecture}. 

\begin{definition}[architecture]
An \emph{\textbf{architecture}} is a complex\footnote{We do not call this complex ``structure'' to avoid a collision with the model-theoretic notion of an $L$-structure over a signature $L$.}
$\mathscr{A}$, written $\langle A, \mathbf{R}, \mathbf{F} \rangle $, 
with
\begin{enumerate}
\item 
$A$, a (possibly empty\footnote{We also include an empty universe for completeness reasons (\textit{cf.} the definition of $\mT_0$ below). In general, however, we are not interested in pathological forms of architectures here.}) set of \emph{\textbf{elements}}\footnote{sometimes also called \emph{artifacts}} 
of the architecture called the \emph{\textbf{universe}} of $\mathscr{A}$,
\item  
$ \mathbf{R} = \{ R_i = (a_1^{(i)},a_2^{(i)},...,a_k^{(i)} )  \subseteq A^k \} $, 
a countable set of relations $R_i$, $i \in \mathbb{N}$ of arbitrary arity $k \in \mathbb{N}$ over the universe $A$, and 
\item
$\mathbf{F} = \{ f_j \: \lvert \: f_j : A_j^{\star} \subseteq A^m \to A, \ j, m \in \mathbb{N} \} $,
a countable ($j \in \mathbb{N}$) set of functions $f_j : A_j^{\star} \to A$. In order to avoid partial functions, we associate to each function $f_j$ its \emph{domain}, $\text{\emph{dom}}(f_j) = A^{\star}_j$.
\end{enumerate}

We write $R^A$ or $f^A$ when we want to highlight that relation $R$ or function $f$ belongs to architecture $\mathscr{A}$ (and not to $\mathscr{B}$).

We also abbreviate $R(a_1,a_2,....,a_n)$ for $(a_1,a_2,...,a_n) \in R \in \mathbf{R}$.
   
We write $\alpha(R) = k$ or $\alpha(f) = k$ to denote the \emph{\textbf{arity}} of a relation $R$ or a function $f$ over $A^k$; that is $R = (a_1,a_2,...,a_{\alpha(R)} ) \subseteq A^{\alpha(R)}$ or $f = f(a_1,a_2,...,a_{\alpha(f)})$.  

\label{d-architecture}
\end{definition}

While the universe $A$ is fairly easy to motivate, one might question the requirement of $\mathbf{R}$ being a \emph{set} of relations $R_i$ instead of a \emph{multiset}. As relations are unnamed (i.e., just defined by their extension), this might prevent our formalization from being able to represent certain architectural structures. Consider an architecture $\mA$ with $A = \{ a, b \}$ and the two ``relations'' $R_{calls}(a,b)$ and $R_{protects}(a,b)$. As dealing with multisets is cumbersome at times, we rather propose to introduce two additional elements $n_1$ and $n_2$ to the universe $A$ and include these explicitly as additional argument in the two relations, like $R_{calls}(n_1,a,b)$ and $R_{protects}(n_2,a,b,)$. If, for whatever reasons, we want to preserve the arity of $R$ we can create ``dummy'' relations like $R_{calls}(n_1,n_1)$ and $R_{protects}(n_2,n_2)$ to distinguish them.\footnote{In the following we assume that such a disambiguation of relations with the same extension has been suitably performed.}

The functions of set $\mathbf{F}$ may serve several purposes: 
\begin{enumerate}
\item Functions may be used to specify the \textbf{attributes} of an element of the architecture \cite{deBoer-2004} (even though relations do suffice for this, using functions for this might be more \emph{efficient} or \emph{economical}\footnote{in the sense of Mach's \emph{Forschungs\"okonomie}}). 

\item Functions may express architectural design \textbf{constraints} on (suitable subsets of) architectural artefacts \cite{Garlan-2003} or \textbf{analysis rules} \cite{Altoyan-2017} like \emph{connectedness}, \emph{reachability}, \emph{self-sufficiency}, \emph{latency} etc.

\item Communications between components (e.g., processes) may be represented by suitable \textbf{``network communication functions''} \cite{Yanchuk-2006, Asha-2020}.

\item In general, functions may also be used to symbolically specify the \textbf{dynamics of an architecture}, e.g., (i) in the form of actions changing the state of the systems, or (ii) for encoding data transformations \cite{deBoer-2004}.

\end{enumerate}

Because we want to focus on the bare minimum of mathematical syntax, we have deliberately not introduced any internal structure of the universe $A$ like differentiating between nodes, software, applications, services, interfaces, requirements, stakeholders, concerns, names, properties, or any other (minuscule or major) element of the architecture.

Users of an architecture (including architects themselves) rarely look at the full architecture comprising all its elements, relations, and properties, but typically refer to and use so-called \emph{views}. In general, a view may be (colloquially) defined as a part of an architecture (description) that addresses a set of related concerns and is addressed to a set of stakeholders.

To formulate this more exactly, we introduce two notions:
\begin{itemize}
\item \textbf{structured views} which are (even though of smaller size) \emph{architectures} in their own right, and
\item \textbf{unstructured views} or just \textbf{views} representing arbitrary slices through an architecture.
\end{itemize}

\begin{definition}[sub-architecture, structured view]		\label{def-sub-arch}
Let 
$\mathscr{A} = \langle A, \mathbf{R}^A, \mathbf{F}^A \rangle $ and 
$\mathscr{B} = \langle B, \mathbf{R}^B, \mathbf{F}^B \rangle $ 
be two architectures. 

Then $\mA$ is called a \emph{\textbf{sub-architecture}} or \emph{\textbf{structured view}} of $\mB$, notation $\mA \subset \mB$, iff
\begin{align}
	A & \subset B \\
	\mathbf{R}^A & \subseteq \{ R^A ~ \lvert ~ R^A = R^B \cap A^{\alpha(R)} \quad \forall 
		R^B \in \mathbf{R}^B \}, \\
	\mathbf{F}^A & \subseteq \{ f^A ~ \lvert ~ f^A = f^B \lvert A  \quad 
												\forall f^B \in \mathbf{F}^B \}.
\end{align}
If $\mA \subset \mB$ we also say that $\mB$ is a \emph{\textbf{super-architecture}} of $\mA$ and write $\mB \supset \mA$.
\end{definition}

The notation $f^A = f \lvert A$ means that $f^A$ is identical to $f$ restricted to the domain $A$. 
Provisions (2) and (3) in the above definition serve to restrict the relations and functions to the subdomain $A \subset B$. 

Dropping the requirements that a view of an architecture needs to be a (sub-) architecture of its own, we define an (unstructured) view as follows:

\begin{definition}[(unstructured) view]
Let 
$\mathscr{A} = \langle A, \mathbf{R}, \mathbf{F} \rangle $
be an architecture. Then the complex $V = \langle A_V, \mathbf{R}_V, \mathbf{F}_V \rangle $
is called an \emph{\textbf{(unstructured) view}} of the architecture $\mA$ iff
\begin{align}
	A_V 			& \subseteq A, \\
	\mathbf{R}_V 	& \subseteq \mathbf{R}, \\
	\mathbf{F}_V	& \subseteq \mathbf{F}.
\end{align}
\end{definition}

This definition deliberately accepts the possibility that one may construct a view $V$ comprising a set $A_V \subset A$ and a relation $R_V \in \mathbf{R}_V$ which are incompatible, i.e., that 
$\lnot R_V(a_1,a_2) ~ \forall a_1, a_2 \in A_V$. One encounters this in situations where just \emph{naming} relation $R_V$ in a view is already conveying important information to one of the stakeholders.

Methodologically, a view is\footnote{or rather: should be} specified by means of a viewpoint, which prescribes the concepts, models, analysis techniques, and visualizations that are provided by the view. Simply put, a view is what you see and a viewpoint is where you are looking from \cite{Archimate}. This is used in the following definition:

\begin{definition}[viewpoint]
Let $\mathscr{A} = \langle A, \mathbf{R}, \mathbf{F} \rangle $ be an architecture and 
\begin{equation}
	V^A = \{ V ~ \lvert ~ V \text{\emph{ is a (structured or unstructured) view of }} \mathscr{A} \}
\end{equation}
be the set of all views of $\mathscr{A}$. Then an injective map
\begin{align*}
	W_I : I & \to V^A \\
	    i & \mapsto V_i^A \in V^A
\end{align*}
is called a \emph{\textbf{viewpoint}}.
\end{definition}

In our definition, a viewpoint may comprise many arbitrary views (selected by the index set $I$) of both kinds, \emph{structured} or \emph{unstructured} ones. This corresponds to cases where one wants to use several different architectural models for describing some aspects of a system. For \emph{economic} reasons we rather want to have $i \ne j \Rightarrow V_i^A \ne V_j^A$, hence $W_I$ has to be injective.

\subsection{n-tier Architectures}		\label{sec-n-tier-architectures}

Let us now link the abstract concepts and definitions more closely to architectural practice. We first observe that the majority of architectural diagrams consists of a collection of "boxes" or other closed shapes (typically representing components or other architectural artefacts) with suitable "connectors", i.e., lines drawn in the space between the boxes connecting two boxes (representing the association between the two entities) \cite{Bass-2013}. These ``boxes \& connectors'' architectures can be formally defined as follows.

\begin{definition}[boxes \& connectors (B\&C) architecture] \label{def-bnc}
An architecture $\mA = \langle A, \mathbf{R} \subseteq \mathcal{P}(A^2), \emptyset \rangle $ is called a \emph{\textbf{boxes \& connectors (B\&C) architecture}}.

We shall often write $\mA = \langle A, \mathbf{R} \rangle $ for a B\&C architecture in the following.
\end{definition}

In passing, we note that not all $n$-ary relations may be represented by binary relations without loss. The expressiveness of B\&C architectures, thus, is limited.

Tiered\footnote{Trivially, the same concept can be used to denote and define \emph{layered} architectures as our definition is ignorant of the actual graphical or pictorial representation (or ``architectural diagram'') used to represent it to our visual sense.} architectures paradigmatically capture the fundamental (not only software) engineering design principle of \emph{separation of concerns}. A tier thereby hides the implementation or execution complexity of one adjacent tier while providing services to the other adjacent tier. This not only enormously aids architects and developers alike, but may also be used for other tasks, e.g., optimization \cite{Chiang-2020}.

In full abstraction, $n$-tier architectures can then be rigorously defined as follows.

\begin{definition}[$n$-tier architecture]
\label{def-n-tier-arch}
Let $\mA = \langle A, \mathbf{R}, \mathbf{F} \rangle $ be an architecture.
Let $C = \{ C_1, C_2, ...., C_n \} \subset \mathcal{P}(A)$ be a \emph{partition} of $A$ into $n$ disjunct sets\footnote{i.e., $\bigcup^n_{i=1} C_i = A$ and $C_i \cap C_j = \emptyset \quad \forall i \ne j$.}. 
Let $k = \alpha(R)$ denote the arity of relation $R \in \mathbf{R}$. 

Then $\mA$ is an \textbf{$\mathbf{n}$-\emph{tier architecture}} iff $\forall R \in \mathbf{R}$ and $\forall a_1,...,a_k \in A$
\begin{equation*}
	R(a_1,...,a_j,..., a_k) \ \Rightarrow 	\ 
	\exists \: l, 1 \le l < n: \ a_j \in C_l \ \vee \ a_j \in C_{l+1} \ 
									\forall \, j=1,...,k.
\end{equation*}
In this case, the classes $C_i$ are called the \emph{\textbf{tiers}} of the architecture. 
An element $a \in C_i$ is said to \emph{\textbf{belong}} to tier $i$.
Tiers $i, j$ with $| i - j | = 1, i \ne j$ are called \emph{\textbf{adjacent}}.
\end{definition}

Informally, in an $n$-tier architecture any element $a$ in tier $i$ (i.e., $a \in C_i$), has only relations with either another element in the same tier or in an adjacent tier $i-1$ or $i+1$. Furthermore, relations never span more than two tiers; they always contain either elements of a single tier only or just of two adjacent tiers. We note that both peer-to-peer (P2P) and client/server (C/S) architectures are 2-tier architectures in our sense. This is expected in and consistent with a purely syntactical approach that (by design) cannot distinguish between the ``roles'' of the two entities in a 2-tier setting. Only semantics (e.g., of the relations connecting the two tiers) is able to provide this discrimination.

We note in passing that every $n$-tier architecture may be turned into an ($n-1$)-tier architecture by simply merging two adjacent tiers $C_i$ and $C_{i+1}$ into a new tier $C_i^\star$. While this might be regarded as annoying from a methodological (or foundational) point of view, it actually is encountered frequently in (real, corporate) architectural work where architecture (diagrams) is (are) often simplified in this way. One option to partially remedy this situation is to define a \emph{maximal $n$-tier architecture}, which is an $n$-tier architecture that cannot be arranged into an ($n+1$)-tier architecture.

We will now introduce some special architectures needed later. The empty architecture $\mT_0$ is introduced for the sake of completeness and closure and may become important in the later development of this line of thinking. The symbol $\mT$ will be motivated later. 

\begin{definition}[empty architecture]
The architecture 
\begin{equation*}
	\mT_0 = \langle \, \emptyset, \emptyset, \emptyset \, \rangle 
\end{equation*}
is called the \emph{\textbf{empty architecture}}.
\end{definition}

We call the next more complex architecture the \emph{trivial architecture} $\mT_1$. Admittedly, it does not bring too much structure with it, but exhibits a slight metamathematical twist, though:

\begin{definition}[trivial architecture]
Let $\text{\emph{id}}_A : A \to A, a \mapsto \text{id}_A(a) = a$, be the identity function on $A$.
Then the architecture 
\begin{equation*}
	\mT_1 = \left\langle A = \left\{ a \}, \{ \, \{ (a,a) \} \, \right\}, \{ \text{\emph{id}}_A \}
			\right\rangle 
\end{equation*}
is called the \emph{\textbf{trivial architecture}}.
\label{d-trivial-arch}
\end{definition}

As the universe $A$ of $\mT_1$ only contains a single element, no attributes or properties of the architecture are indicated at all. Consequently, we also do not get to know any structural properties of this architecture other then identity ($a = a$) and idempotence ($\text{id}_A$). Note in passing that $\mT_0 \subset \mT_1$.

We have introduced the generic $n$-tier architectures above. As we shall later see, they are closely related to an elementary form $\mT_n$, defined as follows.

\begin{definition}[elementary $n$-tier architecture]		\label{def-elementary-n-tier-arch}
Let $T_n = \{ 1, 2, ..., n \}$\footnote{We admit a slight, albeit irrelevant, notational imprecision at this point: In definition \ref{d-trivial-arch} we have denoted the single element of the universe $a$, while here we simply equate the $n$ elements with the first $n$ integers. As we later shall see (in definition \ref{def-homomorphism}) these conventions are isomorphic and, hence, irrelevant when it comes to discussing the structural properties of our architectures. They do, though, simplify our proofs.} 
and define the symmetric binary relation $T^L$ ("linked to") over $T_n \times T_n$, i.e., $T^L(i,j)
\; \forall \, i,j \in T_n$ as follows: 
\begin{align*}
	| \, i - j \, | \ \le 1 	& \Rightarrow \ T^L(i,j), \\
	| \, i - j \, | \ > 1		& \Rightarrow \ \lnot T^L(i,j).
\end{align*}
Then the B\&C architecture
\begin{equation*}
	\mT_n = \langle \: T_n , \: \{ \: T^L \: \} \: \rangle 
\end{equation*}
is called the \emph{\textbf{elementary $n$-tier architecture}}.

\end{definition}

Obviously, when we want to construct an elementary $n$-tier architecture we cannot go below or beyond $n$ elements in the universe $A$ of the architecture. And relation $T^L$ on the set of the first $n$ integers is a very primitive relation on $T_n \times T_n$.\footnote{It is so primitive that it can almost immediately be read off the Peano axioms.}

\subsection{Architecture Homomorphisms}
Having defined the syntax of architectures we now turn to formalizing concepts for comparing different architectures in order to determine whether they are \emph{similar} or even \emph{equal} to each other. This task is typically accomplished by defining suitable structure-preserving maps from one architecture to another, so-called \emph{(homo-)morphisms}.

\begin{definition}[homomorphism, isomorphism]
Let $\mA = \langle A, \mathbf{R}^A, \mathbf{F}^A \rangle $ and 
$ \mB = \langle B, \mathbf{R}^B, \mathbf{F}^B \rangle $
be architectures. Then a (generalized) map $h : \mA \mapsto \mB $, consisting 
of three individual maps, $h = \langle h_0, h_R, h_F \rangle$, 
is called a \emph{\textbf{homomorphism}} between $\mA$ and $\mB$, and $\mA$ is said to be \emph{homomorphic} to $\mB$ iff
\begin{align*}
		h_0 : A & \to B,	\\
		      a & \mapsto h_0(a) \in B. 		\\[5pt]
		h_R : \mathbf{R}^A	& \to \mathbf{R}^B,	\\
		      R^A 			& \mapsto h_R(R^A)  
		      						\qquad \text{\emph{ with }} \alpha(R^A) = \alpha(R^B). \\[5pt]
		h_F : \mathbf{F}^A 	& \to \mathbf{F}^B, \\
			  F^A			& \mapsto h_F(F^A) \qquad \text{\emph{ with }} \alpha(F^A)= \alpha(F^B).
\end{align*}
(with $\alpha(\cdot)$ denoting the \emph{arity} of a relation $R$ or function $f$. See definition \ref{d-architecture}) and
\begin{align}
	R^A(a_1,a_2,....,a_m)	\ 	& \Rightarrow 
				\quad h_R(R^A) \left( \, h_0(a_1), h_0(a_2),..., h_0(a_m) \, 																										\right), \label{equ-hom-r}	 \\  
	h_0 \left( f^A( a_1,a_2,...,a_k) \right) \ & = 
					\quad h_F(f^A) \left( \, h_0(a_1), h_0(a_2),...,h_0(a_k) \, \right).
																	\label{equ-hom-f}
\end{align}
$h$ is called \emph{\textbf{injective}, (\textbf{surjective},
\textbf{bijective})} if all three maps $h_0$, $h_R$, and $h_F$ are injective (surjective, bijective) on their respective (co-)domains. 

In case $h$ is bijective and (\ref{equ-hom-r}) is valid in both directions, i.e.,
\begin{equation}
	R^A(a_1,a_2,....,a_m)	\  \Leftrightarrow \quad 
						h_R(R^A)( \, h_0(a_1), h_0(a_2),..., h_0(a_m) \, )  
\label{equ-isomorphism-r}						
\end{equation}
h is called an \emph{\textbf{isomorphism}} between $\mA$ and $\mB$.
\label{def-homomorphism}
\end{definition}

Condition \ref{equ-isomorphism-r} is required to render our definition of architecture isomorphism equivalent to the (generalized) notion well known from category theory \cite{Barr-1989, Leinster-2014}. There a morphism $f: A \to B$ is an isomorphism iff there exists a morphism $g: B \to A$ such that $ f \circ g = \text{id}_B$ and $ g \circ f = \text{id}_A$.

The following corollary finds that a set of architectures endowed with homomorphisms as per def.~\ref{def-homomorphism} constitute a \emph{category} (which we shall call \textbf{Arch}) \cite{Leinster-2014}.

\begin{corollary}
Let $\mathbb{A} = \{ \mA_1, \mA_2,... \}$ be a set of architectures and 
$\mathbb{H}_A = \text{\emph{Hom}}_A$ be the set of homomorphisms over $\mathbb{A}$.

Then the pair $\langle \, \mathbb{A}, \mathbb{H}_A \, \rangle$ forms a \emph{\textbf{category}} (which we shall call \emph{\textbf{Arch}}).

\end{corollary}
\begin{proof}
We need to show that the ``arrows'' $f \in \mathbb{H}_A$, that is, the architecture homomorphisms as defined above, allow (i) an associative composition ($f \circ g$) and (ii) have an identity element $\text{id}^\mA$ for all $\mA \in \mathbb{A}$.
Let's define $h = g \circ f$ as follows:
\begin{equation*}
\begin{array}{rcl}
	h_0 & = & g_0 \circ f_0,	\\
	h_R & = & g_R \circ f_R,	\\
	h_F & = & g_F \circ f_F.
\end{array}
\end{equation*}
It is clear, that this definition automatically fulfils associativity but we still need to show that our $h$ fulfils the additional constraints (\ref{equ-hom-r}) and (\ref{equ-hom-f}). For the relational part $h_R$ of $h$ we find for any $R \in \mA$\footnote{shorthand for $R \in \mathbf{R}^A$ with 
$\mA = \langle A, \mathbf{R}^A, \mathbf{F}^A \rangle$.}
\begin{equation*}
\begin{array}{rcl}
	R^A(a_1,...) & \Rightarrow & f_R(R^A)(f_0(a_1),...)		\\[3pt]
				 & \Rightarrow & g_R \left( f_R ( R^A ) \right) ( g_0( f_0 (a_1) ),... ) \\[3pt]
				 & =		   & (g_R \circ f_R) (R^A) \left( (g_0 \circ f_0)(a_1,... \right) \\[3pt]
				 & =		   & h_R ( R^A ) ( h_0 (a_1),... ).
\end{array}
\end{equation*}
For the functional part $h_F$ of $h$ we find for any $\phi^A \in \mA$:
\begin{equation*}
\begin{array}{rcl}
	h_0( \phi^A (a_1,...) ) & = & (g_0 \circ f_0) ( \phi^A (a_1),... )  	\\[3pt]
		& = & g_0 \big( \, f_0 \big( \phi^A ( a_1 ),...\big) \big) 			\\[3pt]
		& = & g_0 \big( \, f_F (\phi^A) ( f_0(a_1),... \big)				\\[3pt]
		& = & g_F  ( f_F ( \phi^A) ) ( g_0 ( f_0 (a_1)),... )  				\\[3pt]
		& = & (g_F \circ f_F)(\phi^A) \big( (g_0 \circ f_0)(a_1),... \big)	\\[3pt]
		& = & h_F (\phi^A) ( h_0 (a_1),...).
\end{array}
\end{equation*}
The identity arrow $\text{id}^\mA : \mA \to \mA$ for every object $\mA \in \mathbb{A}$ may be easily defined as 
\begin{equation*}
 \text{id}^\mA = \langle \, \text{id}_A, \text{id}_{\mathbf{R}^A}, \text{id}_{\mathbf{F}^A} \, \rangle,
\end{equation*}
that is the identity map on the respective three domains $A$, $\mathbf{R}^A$, and $\mathbf{F}^A$. This automatically ensures the required identity law, that $\forall f: \mA \to \mB$ we have
	$f \circ \text{id}_{\mA} = f = \text{id}_{\mB} \circ f$, which concludes the proof that the pair
$\langle \, \mathbb{A}, \mathbb{H}_A \, \rangle$ indeed forms a category, \textbf{Arch}.
\end{proof}

We have defined the empty architecture $\mT_0$ and the trivial architecture $\mT_1$ earlier. In categorical terms, these special elements are \emph{initial} and \emph{terminal} objects as the following corollary shows.

\begin{corollary}[Initial \& terminal objects of \textbf{Arch}]
The empty architecture $\mT_0$ is the \textbf{\emph{initial}} object in category \textbf{Arch}.

Let \textbf{ArchBC} be the subcategory of \textbf{Arch} where all architectures $\mA \in 
\textbf{\emph{ArchBC}}$ are restricted to be B\&C architectures.
Then the trivial architecture $\mT_1$ is the \textbf{\emph{terminal}} object in subcategory \textbf{ArchBC}.
\end{corollary}
\begin{proof}
Recall that an object $\mI \in \textbf{Arch}$ is called an \emph{initial} object, if for every element $\mA \in \textbf{Arch}$ there exists a \emph{unique} homomorphism $h: \mI \to \mA$. An element $\mT$ is called \emph{terminal} object, if for every $\mB \in \textbf{Arch}$ there exists a unique homomorphism $g: \mB \to \mT$.

The first claim that $\mT_0 = \langle \emptyset, \emptyset, \emptyset \rangle$ is initial in \textbf{Arch} is trivial. Because all initial object are isomorphic to each other, we may speak of \emph{the} initial object (and also \emph{vice versa} of \emph{the} terminal object).

Regarding the second claim that $\mT_1 = \langle \, T_1 = \{ a \}, \{ \, \{ (a,a) \} \, \}, 
\{ \text{id}_{T_1} \} \, \rangle$ we need to observe that as per our definition \ref{def-homomorphism} homomorphisms $f: \mB \to \mT$ do not change the \emph{arity} of any relation $R^B \in \textbf{R}^B$ or function $g: T_1^n \to T \ \in \textbf{F}^B$. As our (proposed) terminal object $\mT_1$ only contains a single binary relation, $\Delta^2_a = \{ (a,a) \}$, and a single unary function, $\text{id}_{T_1}: T_1 \to T_1, a \mapsto a$, we (i) either have to restrict the set of allowed architectures for which $\mT_1$ should act as terminal, or (ii) we have to generalize the definition of $\mT_1$, or (iii) we change our definition \ref{def-homomorphism} of homomorphisms on \textbf{Arch}.

If we simply restrict the domain of $g$ to B\&C architectures 
$\mB = \langle B, \textbf{R}^B \subseteq \mathcal{P}(A \times A), \emptyset \rangle$ then it is easy to explicitly write down the unique $g: \mB \to \mT_1$:
	\begin{align*}
			g_0 : B & \to T_1 = \{ a \},	\\
		      b & \mapsto a \quad \forall \, b \in B, 			\\[4pt]  
		g_R : \mathbf{R}^B	& \to \{ \Delta^2_a	\} \\
		      R^B 			& \mapsto \Delta^2_a  
		\end{align*}

The uniqueness of $g$ (i.e., its elements $g_0$ and $g_R$) is easy to observe because there simple is only a single way to map every object of $\text{dom}(g)$ to a single element in the codomain 
$\text{cod}(g)$.
We also do not need to specify a $g_F: \mathbf{F}^B \to \{ \, \text{id}_{T_1} \}$ 
because B\&C architectures, by definition, do not contain any functions. Hence, $\mT_1$ is terminal in \textbf{ArchBC}.
\end{proof}

It is somewhat enlightening to investigate how we would have to expand the definition of $\mT_1$ if we want to make this new $\mT^\star_1$ terminal in the full category \textbf{Arch}. Our strategy is based on the observation above that we need to endow $\mT^\star_1$ with suitable relations and functions over $T_1$ of arbitrary arity which may serve as the targets for the maps $g: \mB \to \mT^\star_1$. Let's fix the notation first and write
\begin{equation*}
	\Delta^n_a = \{ ( \underbrace{a,a,...,a}_{n} ) \},
\end{equation*}
for the diagonal relation of arity $n$ over $T_1$ and also abbreviate a (countable) set of generalized identity functions $f^{(k)}_a$ of increasing arity
\begin{equation*}
\begin{array}{lrl}
	f^{(1)}_a: 	& T_1 	& \to T_1,							\\
				& a		& \mapsto f^{(1)}_a(a) = a			\\[6pt]
	f^{(2)}_a: 	& T_1^2 	& \to T_1,		\\
				& (a,a) & \mapsto f^{(2)}_a(a,a) = a			\\[6pt]
	f^{(3)}_a: 	& T_1^3	& \to T_1,		\\
				& (a,a,a) & \mapsto f^{(3)}_a(a,a,a) = a		\\[6pt]
	\ \vdots 	& \vdots & \vdots							\\[6pt]
	f^{(k)}_a: 	& T_1^k	& \to T_1,		\\
				& (\underbrace{a,a,...,a}_{k}) & \mapsto f^{(k)}_a(\underbrace{a,a,...,a}_{k}) = a.	\\
		\ \vdots 	& \vdots & \qquad \vdots
\end{array}
\end{equation*}
With these generalizations it is straightforward to write down our terminal object $\mT^\star_1$ of \textbf{Arch}.
\begin{equation}
	\mT^\star_1 = \langle \: T_1, \{ \Delta^2_a, \Delta^3_a, \Delta^4_a,...,\Delta^n_a,... \},
	 				\{ f^{(1)}_a, f^{(2)}_a, f^{(3)}_a,..., f^{(k)}_a,... \} \, \rangle.
\end{equation}
The unique $g: \mB \to \mT_1^\star$ then may be defined as follows:
\begin{align*}
	g_0: B 				& \to T_1,		\\
		 b 				& \mapsto a \quad \forall \, b \in B,		\\[5pt]
	g_R: \mathbf{R}^B	& \to \{ \Delta^2_a, \Delta^3_a,...,\Delta^n_a,... \},		\\
		 R^B			& \mapsto \Delta^n_a	\qquad \text{if } \alpha(R^B) = n,	\\[5pt]
	g_F: \mathbf{F}^B   & \to \{ f^{(1)}_a, f^{(2)}_a,..., f^{(k)}_a,... \},	\\
		 f^B			& \mapsto f^{(k)}_a		\qquad \text{if } \alpha(f^B) = k.	
\end{align*}

\subsection{Modules and Modularity}		\label{sec-modules-and-modularity}

We have introduced the concept of $n$-tier architectures as one way to realize architectural \emph{separation of concerns} earlier. In systems engineering and software engineering in particular and in information technology in general, we regularly encounter the segmentation of larger systems into distinct \emph{modules}---called \emph{modularization}---to achieve this on a finer scale \cite{Efatmaneshnik-2020}. 

Contrary to the approach in section \ref{sec-n-tier-architectures} we will start the formalization with the elementary form of a modular architecture by suitably generalizing our notion of an elementary $n$-tier architecture (cf.\ definition \ref{def-elementary-n-tier-arch}).\footnote{As we shall see later in section \ref{sec-modularity-at-syntax-level} we do not really have another option.}

\begin{definition}[Elementary modular architecture of $N$ modules]	\label{def-elementary-modular-arch}
Let $M_N = \{ m_1, m_2,...,m_N \}$ be a finite set and 
$\mM_N = \langle \, M_N, \{ D^M \}, \emptyset \, \rangle$ a B\&C architecture with a 
single relation $D^M \subseteq M_N \times M_N$ (with intended semantics of \emph{depends on}).

Then $\mM_N$ is called an \emph{\textbf{elementary modular architecture of $N$ modules}}.

The elements $m_i \in M_N$ are called \emph{\textbf{modules}} of $\mM_N$.
\end{definition}

The following corollary \ref{cor-el-n-tier-is-modular} provides proof that, indeed, definition 
\ref{def-elementary-modular-arch} generalizes the concept of an $n$-tier architecture

\begin{corollary}		\label{cor-el-n-tier-is-modular}
Every elementary $n$-tier architecture $\mT_n$ is an elementary modular architecture $\mM_n$ with $n$ modules. 
\end{corollary}
\begin{proof}
This immediately follows by comparing the definitions of an elementary $n$-tier architecture $\mT_n = \langle \, T_n = \{ 1,2,...,n\}, T^D \subset T_n \times T_n \, \rangle$ with the definition of an elementary modular architecture with $n$ modules, 
$\mM_n = \langle \, M_n = \{ m_1, m_2,...,m_n \}, D^M \subseteq M_n \times M_n \, \rangle$.
\end{proof}

Instead of trying to define a generic modular architecture (as we have done in definition 
\ref{def-n-tier-arch} above) we will invoke our notion of architecture homomorphism as per definition
\ref{def-homomorphism} to define an arbitrary modular architecture comprising $n$ elements (= modules).

\begin{definition}[Modular architecture and modules]		\label{def-modular-architecture}
Let $\mM_n = \langle \, M = \{ m_1, m_2,....,m_n\}, D^M \, \rangle$ be the
elementary modular architecture with $n$ modules.

An (arbitrary) B\&C architecture $\mA = \langle \, A, \mathbf{R}^A \, \rangle$ is called a \emph{\textbf{modular architecture with $n$ modules}} iff there exists a surjective 
homomorphism $h: \mA \to \mM_n$ from $\mA$ to $\mM_n$.

Denote the two maps comprising $h$ as follows: $h = \langle h_0, h_R \rangle$ with
\begin{align*}
	h_0 : A 			& \to M_n,					\\
		  a 			& \mapsto m \in M_n	\quad \forall \, a \in A,	\\[6pt]
	h_R : \mathbf{R}^A 	& \to \{D^M\},				\\
		   R^A 			& \mapsto D^M \quad \forall \, R^A \in \mathbf{R}^A,	
\end{align*}
and define the \emph{pullback} $h^-_0(m)$ of an element $m \in M_n$ as
\begin{equation*}
	h^-(m) = \{ a \in A \ | \ h_0(a) = m \in M_n \}.
\end{equation*}
Then the $n$ pullbacks $A_i = h^-_0(m_i) \subseteq A$ are called the \emph{\textbf{modules}} of $\mA$ with regard to the modularization $\mM$.
\end{definition}
Note that the $A_i$'s form a partition of $A$ into $n$ disjunct (and non-empty) subsets as well. 
The additional characterization of the modularization of $\mA$ ``with regard to $\mM$'' is required because there may be other ways how to segment $\mA$ into $n$ distinct modules with a certain dependency structure (as mediated by the relation $D^M$) between the modules 
(see section \ref{sec-modularity-at-syntax-level} for more details).

\section{Application}
\label{sec-Application}

\subsection{Classical architectural standards and practice}
In this section we present evidence why we believe that the definitions given above make sense.
We do this by ``mapping'' concepts of various architectural metamodels to our architecture definition.
\\

\begin{figure}[!htbp]
\begin{center}
\includegraphics[width=\textwidth]{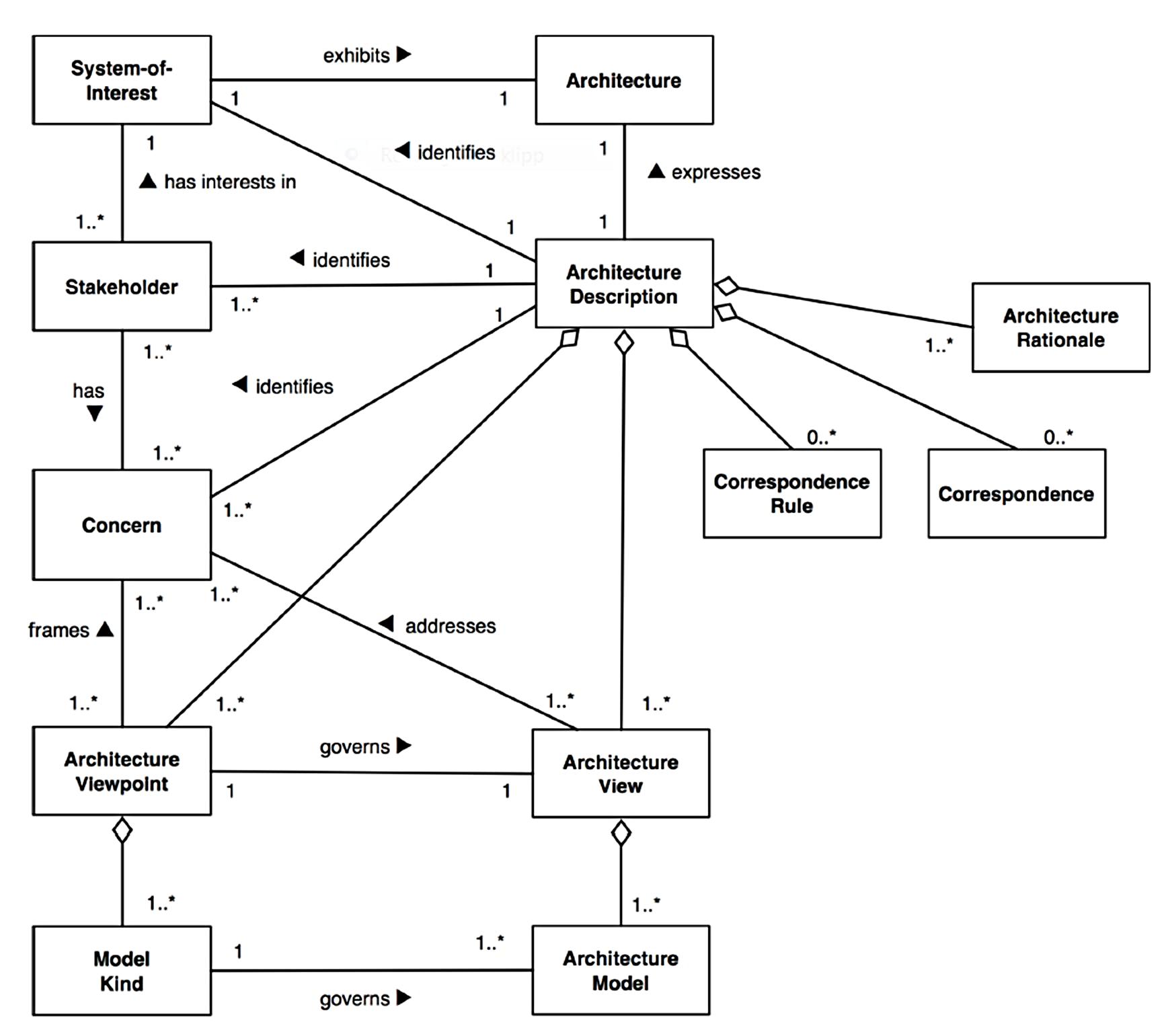}
\end{center}
\caption{ISO/IEC/IEEE 42010: Achitecture metamodel \cite{ISO-42010}}
\label{fig-architecture-metamodel}
\end{figure}

Let us first turn to the ISO/IEC/IEEE Standard 42010:2011 \cite{ISO-42010}. 
Then we can easily identify the elements of the ISO metamodel depicted in 
\ref{fig-architecture-metamodel} which are explicitly formalized by this work. 
This correspondence is depicted in table \ref{t-arch-concepts}. \\

\begin{table}
\begin{tabular}{p{3cm}p{9cm}}
\hline
\textbf{CONSTRUCT} 	& \textbf{RECOGNITION IN THIS THEORY} \\
\hline \hline
\textbf{Architecture} & 
	The theory is capable of expressing architectures for the most complex systems thinkable because the theoretical constructs available basically make up most of our mathematical and logical system.
	\\ \hline

\textbf{Architecture Description} & 
	Our complex $\mA = \langle A, \mathbf{R}, \mathbf{F} \rangle $ essentially conforms
	to an "architecture description". 
	Evidently, it can contain an arbitrary number of elements, relations, and functions
	with essentially unrestricted semantics. 
	\\ \hline
	
\textbf{Stakeholder} &
	Stakeholders may be identified by a suitable subset $S \subset A$.
	Relations over $S$ and supersets $B \supset S$ or functions including elements from $S$ may then 	be used to reason about or include stakeholders in the architecture.
	 \\ \hline
\textbf{Concern} & 
	Concerns may be identified by a suitable subset $C \subset A$.
	Relations over $C$, $S$ and supersets $B' \supset S \cup C $ or 
	functions including elements from $B'$ may then
	be used to reason about or include concerns and their stakeholders in the architecture.
	\\ \hline
	
\textbf{Architecture Viewpoint}  	& 
	Viewpoints $\mathscr{V}$ are an intrinsic element of the theory.
		\\ \hline
		
\textbf{Architecture View} &
	Views $V^A$ are an intrinsic element of the theory.
	 \\ \hline
	 
\textbf{Model Kind} &
	A "model kind" specifies conventions for a certain type of (architectural) modelling. 
	This may be reflected easily in our theory by identifying suitable subsets 
	$K_i \subset (\mathbf{R} \cup \mathbf{F})$ selecting the types of functions and
	relations to be used in a specific "model kind".
	\\ \hline
	
\textbf{Architecture Model} & 
	An architecture view consists of multiple models, each following one model kind.
	Therefore, we can identify an "architecture model" $\mathscr{M}$ by a 
	suitable subset $M \subset A$ and a particular model kind $K_i$.
	\\ \hline
\hline 
\end{tabular}
\caption{Comparison of ISO/IEC/IEEE 42010 and the current theory}
\label{t-arch-concepts}
\end{table}

\begin{figure}[!htbp]
\includegraphics[width=\textwidth]{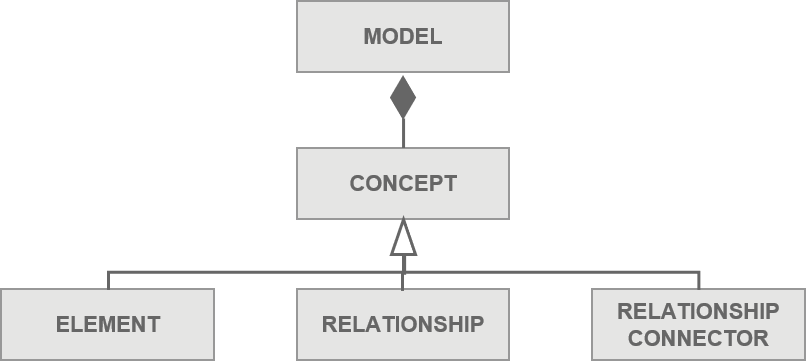}
\caption{Archimate meta-model \cite{Archimate} (own representation)}
\label{fig-archimate-metamodel}
\end{figure}

Figure \ref{fig-archimate-metamodel} shows the metamodel of the \textbf{ArchiMate} \cite{Archimate} framework.
The classes \textsf{ELEMENT} and \textsf{RELATIONSHIP} directly map to our own constructs of def.~\ref{d-architecture}. The class \textsf{RELATIONSHIP CONNECTOR} contains two special elements, \textsf{AND JUNCTION} and \textsf{OR JUNCTION}, which are used to join two or more relations of the same type with the respective semantics of `$\wedge$' and `$\vee$'. The case that \textsf{AND} joins two relations, e.g.~$R(a,b)$ and $R(a,c)$, is easily realized in our formalism as a (new) relation $R(a,b,c)$. In the case of an \textsf{OR}-join, we simply introduce a special element $\omega$ in our universe and can realize the same structure by $R(a,\omega)$, $R(\omega,b)$, and $R(\omega,c)$.

\begin{figure}[!htbp]
\begin{center}
\includegraphics[width=\textwidth]{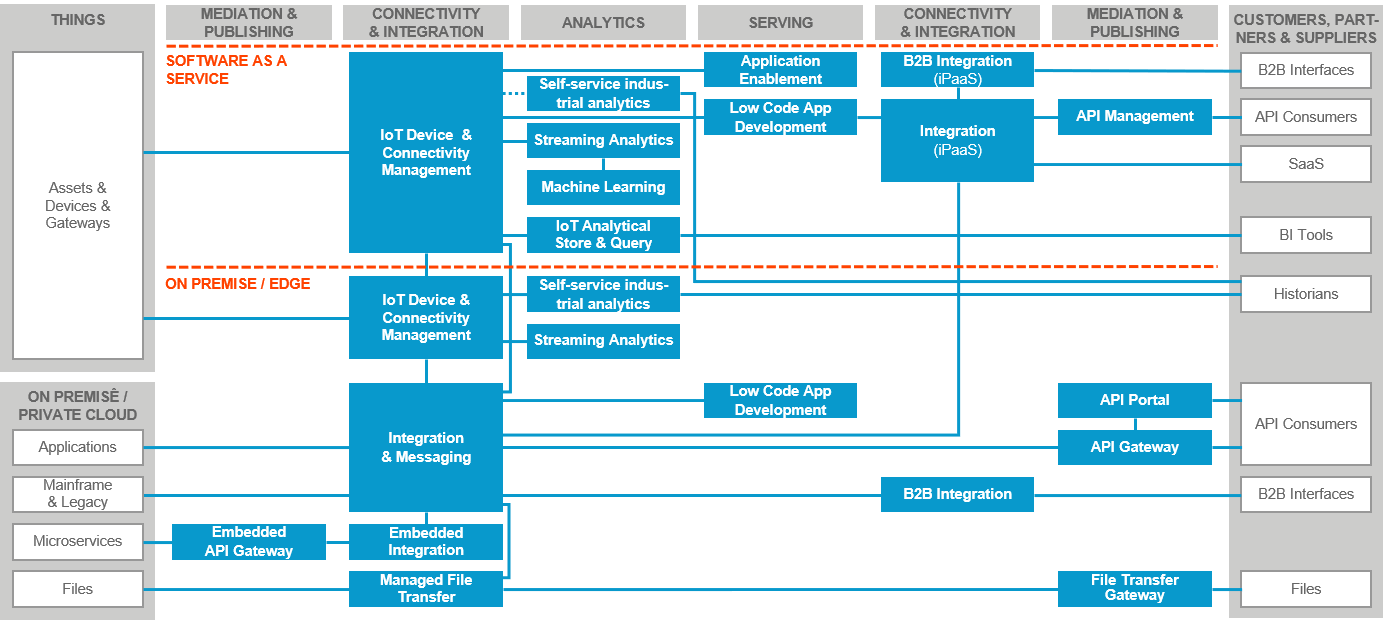}
\end{center}
\caption{Software AG Reference Architecture (Capability View) \cite{SAG-2020}}
\label{fig-sag-refarch}
\end{figure}

Finally, figure \ref{fig-sag-refarch} shows a typical \textbf{boxes \& connectors architecture} prevalent in today's architectural representations and diagrams \cite{SAG-2020}. This particular architecture recognizes a single (binary and symmetric) relation $R(\cdot,\cdot) \subset A^2$ over the universe $A$ of \textsf{capabilities} with the semantic of \textsf{data exchange} or 
\textsf{data flow} between the \textsf{capabilities}.

The architecture (diagram) furthermore differentiates between several vertical \emph{layers}\footnote{but not in the strict sense of an $n$-tier architecture according to def.~\ref{def-n-tier-arch}.} like \textsf{Mediation \& Publishing} or \textsf{Analytics}. There are also two horizontal sections shown in the diagram, 
\textsf{Software as a service} and \textsf{On premise / edge}.

All these constructs may be easily implemented in our rigorous architecture model by specifying certain subsets $L_i \subset A$ (conforming to the various layers and tiers) and defining suitable \emph{characteristic functions} 
$f_{L_i} : A \to \mathbb{B} = \{0, 1 \}$, 
as follows:

\begin{equation*}
	f_{L_i}(a) = 
		\left\{
	 	\begin{array}{cl}
	 		0 & \text{ if } a \in L_i, \\
	 		1 & \text{ if } a \notin L_i.
		\end{array}
		\right.	
\end{equation*}

\subsection{Recent foundational work}

\subsubsection{Foundation of the term \emph{architecture}}

Recently we have seen renewed interest in and attempts at putting concepts, definitions, and terminology in general systems theory (including information technology and engineering) --- such as \emph{architecture}, \emph{structure}, or even the term \emph{system} itself --- on a firmer grounding (cf.~\cite{Wilkinson-2018, Dickersen-2020, Dori-2019}). Here we concentrate on one approach (\cite{Wilkinson-2018, Dickersen-2020}) and demonstrate how our syntactical apparatus is fully capable of formalizing the prose definitions provided in these references.

In this approach, \emph{architecture} is defined as a primary concept and not just one (important) property of the system under study. This definition is, \emph{inter alia}, based on the observation that any architecture invariably includes a description of the relationships of the objects of a system, and that these relationships are further grounded in certain properties that some sets of objects may possess collectively or individually. This is called the \emph{structure} (of a system) and defined as follows:

\begin{definition}		\label{def-wilk-structure}
\textbf{\emph{Structure}} is \emph{junction} and \emph{separation} of the objects of a collection
defined by a \emph{property} of the collection or its objects. 
\end{definition}

On a meta-level, the fact that a system $S$ possesses a certain structure may then be regarded as a (higher level) property of the system. At this (higher) level we can extend our thinking along the lines that also other systems, say $S'$ or $S''$, may possess this property. Such a generalization of a property is called a \emph{type}\footnote{In software engineering, the term \emph{classifier} is also frequently used.}. This then leads to the definition of \emph{architecture} in terms of the property ``having some defined structure'':

\begin{definition}		\label{def-wilk-architecture}
An \textbf{\emph{architecture}} is a \emph{structural type} in conjunction with \emph{consistent properties} that can be implemented in a structure of that type.
\end{definition}

The first example given in \cite{Wilkinson-2018} is a ``system'' of two inhomogeneous linear equations in 2 variables, $x_1$ and $x_2$, the architecture of which is given as the following matrix equation

\begin{equation}	
\mathbf{A}\mathbf{x} = \mathbf{b}
\label{eqn-wilk-ex-1}
\end{equation}
with
\begin{equation}
	\mathbf{A} = 
	\begin{bmatrix}
		a_{11}	&	a_{12} \\
		a_{21}	&	a_{22}
	\end{bmatrix} , \quad
	\mathbf{x} = 
	\begin{pmatrix}
		x_1 	\\
		x_2
	\end{pmatrix},	 \quad
	\mathbf{b} = 
	\begin{pmatrix}
		b_1 	\\
		b_2
	\end{pmatrix}.
	\label{eqn-wilk-ex-2}
\end{equation}

Here the \emph{structural type} consists of (i) a separation of all symbols present in eqn.~\ref{eqn-wilk-ex-2} into three groups, 
$\mathbf{A}$, $\mathbf{b}$, and $\mathbf{x}$, and (ii) two ``junctions''\footnote{cf.~definition \ref{def-wilk-structure} above.}, the juxtaposition of $\mathbf{A}$ and $\mathbf{x}$ and the linking of $\mathbf{x}$ and $\mathbf{b}$ through the equality sign `$=$'.

This can be easily formalized in our mathematical syntax in the following architecture $\mC$:

\begin{equation}
	\mC = \langle \; C = \{ \mathbf{A}, \mathbf{x}, \mathbf{b} \}, \ 
						\{ \, R_{\bullet}, \, R_{=} \, \}, 			
	\ \emptyset \; \rangle.
\end{equation}
with relations $R_{\bullet}$ and $R_{=}$ defined as follows:
\begin{equation}
	R_{\bullet} = \{ \, (\mathbf{A}, \mathbf{x} ) \, \}, \quad 
	R_{=} = \{ \, ( \, \mathbf{x}, \mathbf{b} ) \, \}.
\end{equation}

The \emph{consistent properties} of an architecture (cf.~definition \ref{def-wilk-architecture}) then are the three equalities given in eqn.~\ref{eqn-wilk-ex-2}. We note that these properties are not part of the architecture formalization $\mC$. This feature is also present in the original example in \cite{Wilkinson-2018} and, in our view, a sign of an incomplete formalization. A (syntactically) exhaustive architecture would also have to include these \emph{properties} which may be easily achieved be extending $\mC$ into a super-architecture $\mC^\star$:

\begin{equation}	\label{eqn-wilk-cfs-C}
	\mC^\star = \langle \; \{ C_E^\star, \mathbf{A}, \mathbf{x}, \mathbf{b} \},
				\{ \, R_{\bullet}^\star, \, R_{=}^\star \, \}, \ \{ f^\star \} \; \rangle 
\end{equation}
with $C_E^\star = \{ a_{11}, a_{12}, a_{21}, a_{22}, x_1, x_2, b_1, b_2 \}$, $R_{\bullet}^\star = R_{\bullet}$, and $R_{=}^\star = R_=$. The (new) function $f^\star(\cdot)$ then captures the \emph{property} how the individual variables and constants present in eqn.~\ref{eqn-wilk-ex-2} relate to the structural part of the architecture as given in eqn.~\ref{eqn-wilk-ex-1}:

\begin{equation}
f^\star : C_E^\star \to C \text{ with } \left\{
	\begin{array}{rcll}
		a_{ij} & \mapsto & \mathbf{A}   	& \quad i,j = 1,2;	\\
		x_i    & \mapsto & \mathbf{x}		& \quad i=1,2;		\\
		b_i	   & \mapsto & \mathbf{b}		& \quad i=1,2.
		\end{array}		\right.
\label{eqn-wilk-cfs-f}
\end{equation}
It is easy to verify that the architecture $\mC^\star$ defined in eqn.~\ref{eqn-wilk-cfs-C} is indeed a \emph{super-architecture} of $\mC$, i.e.~$\mC^\star \supset \mC$, in the sense of def.~\ref{def-sub-arch}. Alternatively we may formulate that the matrix equation \ref{eqn-wilk-ex-1} without properties of eqn.~\ref{eqn-wilk-ex-2} only represents a sub-architecture or structured view of the \emph{full} architecture. This also perfectly mirrors the verbal definition of architecture in def.~\ref{def-wilk-architecture} which explicitly requires a ``conjunction'' of the \emph{structural type} with the \emph{consistent properties}.

\begin{figure}[htbp]
\begin{center}
\includegraphics[width=\textwidth]{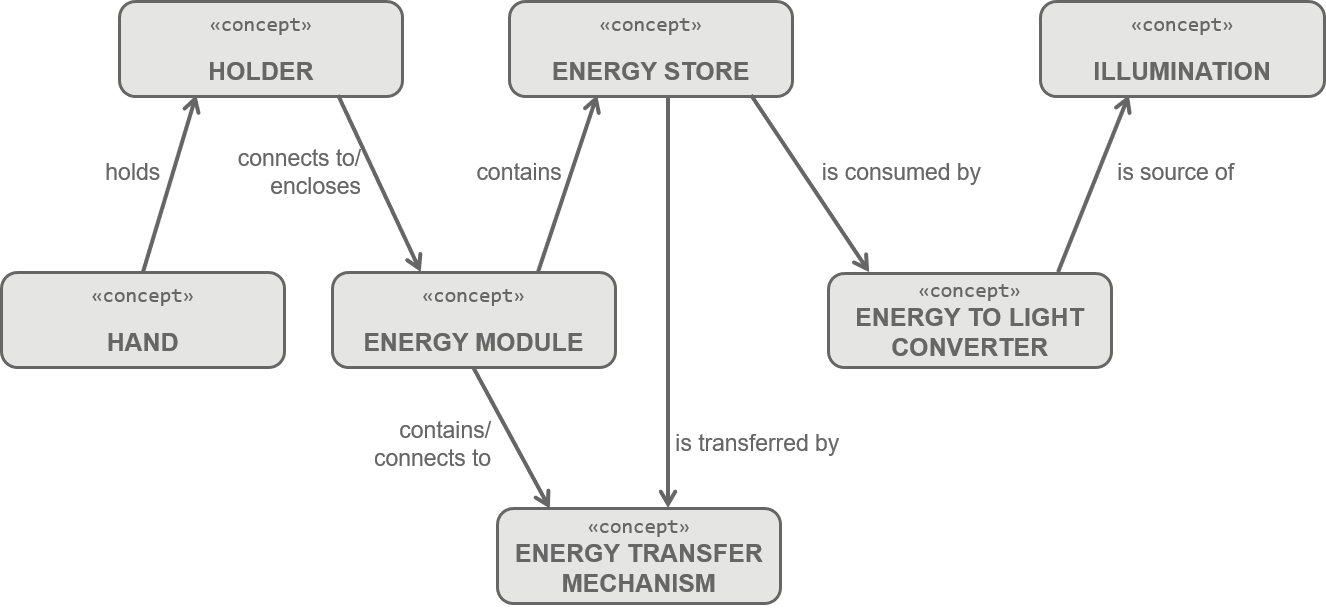}
\end{center}
\caption{Architectural structure of a generic torch (after \cite{Wilkinson-2018})}
\label{fig-wilk-torch}
\end{figure}

In a second example, Wilkinson \textit{et al.}~\cite{Wilkinson-2018} derive the architecture of generic torch-like lighting systems (i.e.~torches using burning wax or batteries) based on the framework of Conceptual Structures \cite{Sowa-1984}. The final architectural structure is given in figure \ref{fig-wilk-torch} with the boxes representing the fundamental ``concepts'' of the architecture and the arrows signifying the relations between them.

It is plainly evident that this architecture is a simple B\&C architecture\footnote{It also happens to be a maximal 6-tier architecture.} with a universe containing 7 elements and 7 binary relations. 

Tangential to our main points we observe that the relation \textsf{is transferred by} between the two concepts \textsf{Energy Store} and \textsf{Energy Transfer Mechanism} does not seem to be correct: Not the \textsf{Energy Store} is transferred but the energy carrier contained in the \textsf{Energy Store} (e.g.~electricity in a battery, wax in a Hessian role). The same imprecision applies to relation \textsf{is consumed by}. This seems to be a consequence of the fact that the architecture is missing the important additional concept of \textsf{Energy} or \textsf{Energy Carrier}.  

This purely relational approach is extended in \cite{Dickersen-2020} where architectures are described using first order predicate calculus restricted to the domain (or \emph{universe} in our terminology) of architectural concepts. As our definition \ref{d-architecture} is sufficiently powerful to be interpreted as a ``model'' (in a model theoretic sense \cite{Doets-1996, Hodges-1993}) of such a ``theory'', the formalization given in this contribution is logically equivalent (by 
G\"odel's Completeness Theorem) to the approach of \cite{Dickersen-2020}. 

\subsubsection{Foundation of the terms \emph{module} and \emph{modularity}}

In the context of system engineering, Efatmaneshnik \emph{et al.}~\cite{Efatmaneshnik-2020} discuss and define the notions \emph{module}, \emph{modularity}, and \emph{modular}. They define a 
system $S$ as a set $S = \{ S_P, S_F, S_{NF} \}$ comprising 
\begin{itemize}
\item the set of (physical) subsystems $S_P$ with $S_P = \{ s_1, s_2, ..., s_{n_P} \}$,
\item the set of functional requirements $S_F$ with $S_F = \{ f_1, f_2, ..., f_{n_F} \}$, and
\item the set of non-functional requirements $S_{NF}$ 
			with \\
			$S_{NF} = \{ r_1, r_2, ..., r_{n_{NF}} \}$.
\end{itemize}
Each of the $n_P$ physical subsystems $s_i$ then consists of several (physical) components, $c_i^j$, vid. $s_i = \{ c_i^1, c_i^2,...., c_i^{n_i} \}$.

Sadly, without any mathematical definition, they introduce the core notion \textbf{\textsf{realizes}} between physical components and the functional and non-functional requirements postulating that physical subsystems \textsf{realize} both, functional and non-functional requirements and write this as follows:
\begin{align*}
		S_P & \xrightarrow{\text{realizes}} S_F,			\\
		S_P & \xrightarrow{\text{realizes}} S_{NF}.
\end{align*}
The notion  \textsf{realizes} is further extended to the level of subsystems as follows:
\begin{equation*}
	s_i \in S_P   \xrightarrow{\text{realizes}} 
						\{f_{i_1},f_{i_2},...,f_{i_m} \} \subseteq S_F,
\end{equation*}
with the meaning that ``a set of main system functions are [\emph{sic!}] satisfied/delivered/realized by a subsystem''.

Let us first fix the imprecise notation from above with the following definition of the ``arrow'' notation before turning to their propositions and claims.

\begin{definition}[arrow notation]
Let $A$ and $B$ be sets. Let $R \subseteq A \times B$ denote a binary relation over 
$A \times B$. Then we write
\begin{align*}
	a \xrightarrow{R} b & \quad \Leftrightarrow \quad (a,b) \in  R, 		\\
	a \xrightarrow{R} \{ b_1,b_2,...,b_k \}\subseteq B	 & \quad \Leftrightarrow \quad
					\forall \: b_i, i=1,...,k : (a,b_i) \in R. \\
\end{align*}
\end{definition}
Then we can a define \emph{system} in the sense of \cite{Efatmaneshnik-2020} precisely as follows.

\begin{definition}[Efatmaneshnik system]	\label{def-ef-system}
A \textbf{\emph{system}} $S$ is a quadruple $S = \langle S_P, S_F, S_{NF}, R \rangle$ comprising the set of (physical) subsystems $S_P$, and the sets of functional ($S_F$) and non-functional ($S_{NF}$) requirements,
\begin{align}
	S_P 	& = \{ s_1, s_2, ..., s_{n_P} \},		\\
	S_F 	& = \{ f_1, f_2, ..., f_{n_F} \},		\\
	S_{NF}	& = \{ r_1, r_2, ..., f_{n_{NF}} \},
\end{align}
and a binary relation $R \subseteq \mathcal{P}(S_P) \times ( S_F \cup S_{NF})$ (with the intended semantics of \emph{\textbf{realization}}), such that
\begin{align}
		S_P & \xrightarrow{\text{R}} S_F,			\\
		S_P & \xrightarrow{\text{R}} S_{NF}.	\label{eq-Sp-arrow_Snf}
\end{align}
Each of the $n_P$ physical subsystems $s_i \in S_P$ may then consist of several (physical) components, $c_i^j$,
\begin{equation}		\label{eq-ef-def-si}
	s_i = \{ c_i^1, c_i^2,...., c_i^{n_i} \},
\end{equation}
fulfilling
\begin{equation}		\label{eq-ef-def-m}
	\{ s_i \}   \xrightarrow{\text{R}} 
						\{f_{i_1},f_{i_2},...,f_{i_m} \} \subseteq S_F,
\end{equation}
for some $f_{i_\alpha}$.
\end{definition}
Note that the left argument of $R(\cdot, \cdot) $ needs to be an element of the powerset of $S_P$, 
$\mathcal{P}(S_P)$, i.e.~itself is a set. We need this to be able to formalize the fact that (at times) two subsystems $s_1$ and $s_2$ collectively realize a functional $f$ or nonfunctional requirement $r$ without any single one of them being able to achieve this. For instance, let $s_i, i=1,2$, denote two servers and let $r_{HA}$ stand for the nonfunctional requirement of \emph{high availability}. Then the fact that an \emph{active/active} configuration of the two servers realizes $r_{HA}$ may be written as
\begin{align*}
	\{ s_1, s_2 \} & \xrightarrow{R} r_{HA},	\\
\intertext{but}
	\lnot \, \big( \, \{ s_1 \} & \xrightarrow{R} r_{HA} \, \big),	\\
	\lnot \, \big( \, \{ s_2 \} & \xrightarrow{R} r_{HA} \, \big).
\end{align*} 

The authors then define \emph{architectural modularity}\footnote{As we will argue later, the name is misleading. This particular form of modularity should be better characterized as ``functional modularity'', not the least to demarcate this notion of \emph{modularity} from the ``non-functional modularity'' the authors introduce and treat in the rest of the paper.} iff $m \equiv 1 \: 
\forall s_i$ in equ.~\ref{eq-ef-def-m}, that is, if it reads

\begin{equation}		\label{eq-ef-one_to_one}
	 \forall \, i = 1,...,n_P : \ \{s_i\} \xrightarrow{R} f_{i_1}.
\end{equation}
They then claim that condition \ref{eq-ef-one_to_one} means ``there is a one-to-one correspondence between subsystems [i.e., the $s_i$'s, CFS] and main functional requirements 
[i.e., the $f_i$'s, CFS]".

\begin{figure}[!htbp]
\begin{center}
\includegraphics[width=\textwidth]{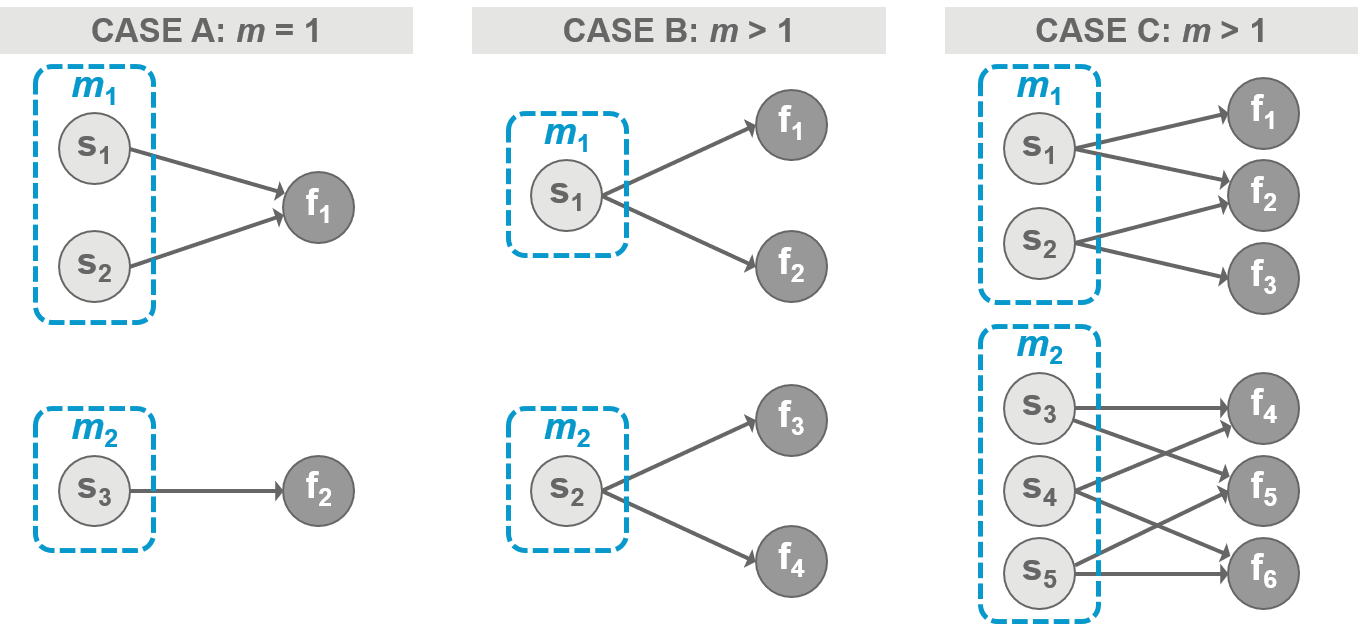}
\end{center}
\caption{Functional modularization}
\label{fig-functional-modul}
\end{figure}

Obviously, this claim is unfounded as the simple system
\begin{equation*}
S^\star = \langle \, \{s_1,s_2,s_3\}, \{f_1,f_2\},\emptyset, R^\star \, \rangle 
\end{equation*}
with relation $R^\star$
\begin{align}
	\{s_1\} & \xrightarrow{R^\star} f_1,	\label{eq-s-1} \\
	\{s_2\} & \xrightarrow{R^\star} f_1,					\\
	\{s_3\} & \xrightarrow{R^\star} f_2,	\label{eq-s-3}
\end{align}
provides a counterexample where we certainly do not have a 1:1 correspondence between the $s_i$'s and the $f_i$'s (see also case A in figure \ref{fig-functional-modul}).

However, the original claim of ``modularity'' may be salvaged if we identify the \emph{modules} $m_k$ not (mistakenly) simply by putting $m_k \equiv s_k$, but by
\begin{equation}		\label{def-modules-m1}
	m_k = \{ s_i \: | \: \{s_i\} \xrightarrow{R^\star} f_k \, \}	\quad k=1,...,n_F.
\end{equation}

The authors furthermore characterize the situation $m > 1$ in condition \ref{eq-ef-def-m}
as an ``architecturally nonmodular" system where ``function sharing'' (according to \cite{Agrawal-2009}\footnote{Ref.~\cite{Agrawal-2009} does not support the notion of ``function sharing'' because it is solely concerned with the \emph{internal} modularization of a product in terms of its constituting components, i.e., of $S_P$ in our syntax, using elements of graph theory and the Design Strategy Matrix method. Neither requirements nor the relation of components or subsystems to them play any role in the core part of that work.}) 
takes place.\footnote{The article mistakenly writes $i_m > 1$ instead of the intended $m > 1$.}

Again, this proposition is not valid universally. Consider the system
\begin{equation*}
S^\star = \langle \, \{s_1,s_2 \}, \{f_1,f_2, f_3, f_4\}, \emptyset, R^\star \, \rangle 
\end{equation*}
with relation $R^\star$
\begin{align}
	\{s_1\} & \xrightarrow{R^\star} \{ f_1, f_2 \},	\label{eq-s-4}	\\
	\{s_2\} & \xrightarrow{R^\star} \{ f_3, f_4 \}.	\label{eq-s-5}
\end{align}
This system $S^\star$ clearly (i.e., for all practical definitions of ``architectural modularity'', in our opinion) consists of two modules, $m_1 = s_1$ and $m_2 = s_2$ (cf.~case B in figure 
\ref{fig-functional-modul}).

In passing we note that figure 2 in ref.~\cite{Efatmaneshnik-2020} is incorrect because the depicted arrows, which are the graphical representation of the relation $\xrightarrow{realizes}$ according to the running text, wrongly extend from the ``logical domain'' (i.e., $S_F$) to the ``physical domain'' (i.e., $S_P$). The figure, thus, is either mislabelled or the direction of all (!) arrows needs to be reversed.

Nevertheless, we may salvage the whole notion of (architectural) modularity by generalizing the observation that in both cases A and B it was possible to identify certain partitions of the set $S_P$ \emph{and} of the set $S_F$ such that there is a 1:1 relationship $\xrightarrow{R^\star}$
between elements of both partitions. Because this segmentation is solely based on the functional requirements $S_F$ we would rather call this a \textbf{\emph{functional modularization}} of $S$ instead of the misleading ``architectural modularization'' as put forward in \cite{Efatmaneshnik-2020}.

This immediately leads to the following universal definition.

\begin{definition}[functional modularization]		\label{def-functional-modularization}
Let $S = \langle S_P, S_F, S_{NF}, R \, \rangle$ be a system.
Let $F = \{ F_1, F_2,..., F_i, ..., F_N \}$ be a partition of $S_F$ into $N$ disjunct subsets $F_i$ and define $R^{-}(\cdot)$ as
\begin{equation}
	R^{-} (F_i) = \{ s \in S_P \: | \: \exists f \in F_i: \{s\} \xrightarrow{R} f \}.
\end{equation}
In case the following condition holds for $F$:
\begin{equation}			\label{def-modules-gen}
	R^{-}(F_i) \, \cap \, R^{-}(F_j) = \emptyset	\quad \forall \, i \neq j, \ i,j=1,....N,
\end{equation}
we call the set $M$
\begin{equation}
	M = \{ m_1,m_2,...,m_N \}, \quad \ m_i = R^{-}(F_i),\; i=1,...,N,
\end{equation}
a \textbf{\emph{functional modularization}} of the system $S$. The $N$ elements $m_i \in M$ are called (functional) \textbf{\emph{modules}} of $S$ (remember $m_i \subseteq S_P$).
\end{definition}

Evidently, the (implicit) partitions of $S_F$ in our previous examples 
(cf.~e\-quations \ref{def-modules-m1}--\ref{eq-s-5}) all fulfil condition \ref{def-modules-gen}. We have depicted a non-trivial example of functional modularization as case C in figure \ref{fig-functional-modul}.

Let us now turn to the central definition of \emph{modularity} as put forward in \cite{Efatmaneshnik-2020}.

\begin{definition}[Efatmaneshnik modularization]		\label{def-ef-module}
A \textbf{\emph{module}} is a group of some of a system’s elements (components,
subsystems, etc.), with a physical or notional boundary, and is detachable, either physically or notionally from the system that, by this detachability alone, has a nonfunctional utility
for one or more system lifecycle stages or stakeholders.

A \textbf{system} is \textbf{\emph{modular}} if it either has modules or has certain
qualities that ease or make possible the process of modularization.

A \textbf{unit} is \textbf{\emph{modular}} if it is either a module itself or is a unit with the potential to be a module.

\textbf{\emph{Modularity}} is an attribute of a system or a unit (of that system) designating (i) the quality of being modular or (ii) the potential to become modular.

For both, a system or a unit of a system, \textbf{\emph{modularization}} is the process of becoming modular.
\end{definition}

At its heart the definition of \emph{module} says that any subset of elements of the system 
satisfying (to any degree) one or more nonfunctional requirements may be regarded a module. 
It is completely irrelevant if this separation is physically possible at all (arg.~``physically detachable'') or not (arg.~``notionally''); the simple ability to discriminate the elements that form part of the module from those which are not part of it suffices.

Consequently, they formalize a module $m$ as follows:
\begin{equation}
		m \subseteq \bigcup^{n_P}_{k=1} s_k, 
\end{equation}
with subsystems $s_k \in S_P = \{ s_1,s_2,...,s_{n_P} \}$. 
Actually, eqn.~5 in \cite{Efatmaneshnik-2020} is imprecise in defining 
$m = \cup c^j_i$ [\emph{sic!}] with the $c^j_i \in s_i$ because the (collective) union operator\footnote{which also should have been written as a big \emph{cup} $\bigcup$ instead of the binary operator $\cup$.} $\bigcup$ is only defined for \emph{sets} and not for individual elements. The correct formulation would have been $m = \bigcup \, \{c^j_i\}$.

The condition in definition \ref{def-ef-module} of ``having a nonfunctional utility'' is modelled by relating $m$ to the set of nonfunctional requirements $S_{NF}$ of the system $S$:
\begin{equation}
	m \xrightarrow{\text{fully or partially realizes}} R_{NF} \subseteq S_{NF}.
\end{equation} 
The introduction of this relation $\xrightarrow{\text{fully or partially realizes}}$ 
is problematic in two ways.

First, this is \textbf{not} the relation $R$ we have introduced in definition \ref{def-ef-system} above but a new one as it ranges over the components $c^j_i$ of the subsystems $s_i$ (cf.~equ.~\ref{eq-ef-def-si} in definition \ref{def-ef-system}).

Secondly, we note that the authors weaken the original meaning of relation $R$ as ``realizing'' (functional requirements) when they now also apparently accept semantics of ``partial fulfilment''.

Ignoring the slightly changed semantics of $R$ we can, nevertheless, salvage the formalization of module in the sense of \cite{Efatmaneshnik-2020} by suitable extending the range of $R$ to a new $R^\star$:
\begin{definition}[Efatmaneshnik modularization 1]		\label{def-ef-module-1}
Let $S$ be a system
\begin{equation*}
	S = \langle S_P = \{ s_1,s_2,...,s_{n_P} \}, S_F, S_{NF}, R^\star \rangle,
\end{equation*} 
with the $s_i = \{ c^1_i, c^2_i,c^3_i,... \}$ its subsystems and components $c^j_i$ thereof.
Abbreviate the set of all components of $S$ by $S_C$, that is
\begin{equation}		\label{eq-def-mod-sc}
		S_C = \bigcup^{n_P}_{k=1} s_k,
\end{equation}
and let $R^\star$ be a binary relation (with the intended meaning of \emph{realizes}) as follows:
\begin{equation}				\label{eq-def-mod-rstar}
	R^\star \subseteq \mathcal{P}\big( S_P \cup S_C \big) \times 
							\big( S_F \cup S_{NF} \big).
\end{equation}
A set $m$
\begin{equation}				\label{eq-def-mod-m}
		m = \{ c^\mu_\alpha, c^\nu_\beta, c^\rho_\gamma,...\} \subseteq S_C 
\end{equation}
is called a \emph{\textbf{module}} iff we have
\begin{equation}			\label{eq-def-mod-m-arrow-R}
	m \xrightarrow{R^\star} R_{NF} = \{ r_{i_1}, r_{i_2},... \} \subseteq S_{NF}.
\end{equation}
for some set $R_{NF}$. A subset $m^\star \subseteq m$ is called a \emph{\textbf{submodule}} of $m$ iff
\begin{equation}
		m^\star \xrightarrow{R^\star} r_i \in R_{NF}.
\end{equation}
\end{definition}

Definition \ref{def-ef-module-1} formalizes the notion of ``detachability'' (cf.~definition \ref{def-ef-system}) of a module from the rest of the system in eqn.~\ref{eq-def-mod-m} in the form of \emph{naming} only.  Eqn.~\ref{eq-def-mod-m-arrow-R} captures the idea that any module $m$ has to have ``nonfunctional utility''.

Unfortunately, the definition above is apparently limited to defining a single module in complete ignorance of any other module the system may possess. This, however, is contrary to architectural or engineering practice where one is not so much concerned with identifying a single module within a system but in sensibly segmenting the \emph{whole} system into a set of (distinct) modules. The authors briefly mention this but do not regard it in any way problematic.

Secondly (and because of the previous shortcoming) the above construction allows that two or more modules arbitrarily share \emph{the very same} components. The authors remark
\begin{quotation}
[...] some define the usage of standard components as modularity.
This means that components $c^j_i$ and $c^k_l$ are the same type of components or are exactly the same component (which can happen in software where a piece of code is called/used by different parts of the program).
\end{quotation}
Now, while their formalism certainly allows this (vid.\ eqn.~\ref{eq-def-mod-m} does not forbid two modules $m$ and $m^\star$ sharing one or more components $c^\sigma_\delta$), this is typically and certainly not the case in real practice. If we have physically detachable modules $m$ and $m^\star$ sharing one or more components it would lead to the considerably strange situation where one can extract module $m$ but, in doing so, destroys module $m^\star$ (or at least depriving it from the shared components).

The incomplete understanding is also evidenced in the quote above where relation \textsf{calling} (e.g.~a subroutine, a library, or another module) is confounded with the notion of \textsf{being member of a module}. Remarkably, none of the three examples of (physical) modularization displays this (artificial) feature.

From a principal point of view, the paper repeatedly argues that every subsystem $s_i$ is a module. but not every module $m$ necessarily is a subsystem. Formally, that translates to the claim
\begin{equation}		\label{eq-every-subsystem-is-a-module}
\forall s_i \in S_P: \exists \, R_i \subseteq S_{NF} : \ s_i \xrightarrow{R^*} R_i.
\end{equation}
While on an informal level we certainly could not agree more with this assertion, the claim itself cannot be (logically correctly) deduced from the formalization of \emph{module} given in the paper or reconstructed here. Technically, this lies in the fact that the two definitions given, eqn.s~\ref{eq-def-mod-m-arrow-R} and \ref{eq-Sp-arrow_Snf}, are too generic to allow a cogent deduction of 
eqn.~\ref{eq-every-subsystem-is-a-module}.

The root cause, though, is the fact that one cannot formalize modularity or define modules on the level of syntax only. This is rigorously proven in our ``no go'' theorem \ref{th-generic-modular-architectures} below. 

The only thing we need to show is that we can apply our ``no go'' theorem to the definitions of 
\cite{Efatmaneshnik-2020}, that is, that the formalization given here in this section is compatible with our own theory (as given in section \ref{sec-Theory}.

For this we first extend definition \ref{def-ef-module-1} to a segmentation of the system $S$ 
into a disjunct set of $N$ modules.

\begin{definition}[Efatmaneshnik modularization 2]		\label{def-ef-module-2}
Let $S$ be a system, $S= \langle S_P = \{ s_1,s_2,...,s_{n_P} \}, S_F, S_{NF}, R \, \rangle$. 
Write $S_C = \bigcup^{n_P}_{i=1} c^j_i$ for the set of all components of $S$ and let 
$M = \{ m_1, m_2,\ldots,m_N \} \subseteq \mathcal{P}(S_C)$ be a partition of $S_C$ 
into $N$ distinct subsets $m_i \subseteq S_C$ such that
\begin{equation}
	\forall i=1,\ldots,N : \exists R_i \subseteq S_{NF} : \ m_i \xrightarrow{R} R_i.
\end{equation}
Then $M$ is called a \emph{\textbf{nonfunctional modularization}} of $S$ and the $N$ $m_i$'s are called \emph{\textbf{modules}} of $S$.
\end{definition}

If we now collect all elements of $S$ in a single (flat) set $S^\star = S_P \cup S_C \cup S_F \cup
S_{NF} \cup M$ and remember that $m_i \xrightarrow{R} R_i$ is just another way to write 
$R(m_i, r_\alpha) \ \forall r_\alpha \in R_i$, then we can write down the B\&C \emph{architecture} $\mathscr{S}$ of the system $S$ as $\mathscr{S} = \langle S^\star, R \, \rangle$
with a suitable $R \subseteq S^\star \times S^\star$.

\subsection{Mathematical Reasoning}
Here we demonstrate that our theory of architectures also provides a mathematical framework for reasoning about (and proving) facts about architectures in general.

\subsubsection{Graphs and Architectures}

\begin{theorem}
Every Graph $G = (V, E)$ with $V = \{ V_1, V_2,...,V_n \} $ the set of vertices and $E \subseteq V^2$ the set of (directed) edges of $G$, is isomorphic (in a model theoretic sense) to a B\&C architecture $\mathscr{G}$.
\end{theorem}

\begin{proof}
Define a B\&C architecture $\mathscr{G}$ as follows:
$\mathscr{G} = \langle V^G = V, \mathbf{R}^G = \{ E \}, 
		\mathbf{F}^G = \emptyset  \rangle $.
Then, in a naive sense, we already can \emph{see} the isomorphism. On an exact level, though, we first have to specify a framework within which we may relate the two objects $G$ and $\mathscr{G}$ to each other. We will use model theory here \cite{Doets-1996, Hodges-1993}. Then graph $G$ and architecture $\mathscr{G}$ may be regarded as $L$-structures of the (simple) signature $L = \{ E \}$ containing just the symbol for the edge relation. 
In this case, two $L$-structures are isomorphic iff there exists a bijective homomorphism $h: V \to V^G$
satisfying the following condition on $E$
\begin{equation*}
	(a,b) \in E \, \Leftrightarrow \, ((h(a),h(b)) \in E \in V^G.
\end{equation*}
The following surjective embedding $h$
\begin{align*}
	h : V 		& \to V^G	\\
		v 		& \mapsto v \in V^G
\end{align*}
provides the (trivial) structure preserving bijection between the two domains $E$ and $V^G$. 
\end{proof}

\subsubsection{Characterization of n-tier architectures}

Many architects use $n$-tier architectures due to their clear (and easy to follow) structure. The following lemma and theorem provide a rigorous characterization in terms of elementary $n$-tier architectures $\mT_n$.

\begin{lemma}
A $n$-tier B\&C architecture $\mA$ is homomorphic to the elementary $n$-tier architecture $\mT_n$, that is there exists a homomorphism $h : \mA \mapsto \mT_n$.
\label{lem-n-tier}
\end{lemma}

\begin{proof}
Let $\mA = \langle A, \mathbf{R} \rangle $ be an $n$-tier B\&C architecture. Then, by definition, every element $a \in A$ belongs to exactly one tier $C_i$ with $ 1 \le i \le n$. 
We also have $R \subseteq A^2$ because of the "boxes \& connectors" property. 
Then define the map $h : \mA \mapsto \mT_n$ as follows:
\begin{equation*}
\begin{array}{lrl}
	h_0: &		A			& \to T_n = \{1,2,....,n\}	\\
	     &       a \in C_i   & \mapsto i \in T_n		\\ \\
	    
	h_R: &		\mathbf{R}	& \to \{\, T^L \, \} 	\\
		 &		R			& \mapsto T^L 
\end{array}
\end{equation*}
Assume $R(c_i,c_j)$ with $c_i \in C_i$ and $c_j \in C_j$. Then $h(c_i)=i$ and 
$h(c_j)=j$. Note that, from now on, we drop the subscripts on the three different maps $h_0$, $h_R$, and $h_F$ comprising $h$ for simplicity and readability reasons. It should be immediately evident from the respective domain $\text{dom}(h_\alpha)$ which $h_\alpha$ we are talking about. 
Because $\mA$ is an $n$-tier architecture, we have 
\begin{equation*}
 | i-j | \le 1 \Rightarrow |\, h(c_i) - h(c_j) \, | \le 1 = T^L ( h(c_i), h(c_j) )
 	= h(R)(h(c_i),h(c_j)).
\end{equation*}
Therefore $h$ is a homomorphism.
\end{proof}

We can strengthen the above lemma to characterize $n$-tier B\&C architectures as follows:

\begin{theorem}
Every B\&C architecture $\mA = \langle A, \mathbf{R} \rangle$ is an $n$-tier B\&C architecture iff there exists a surjective homomorphism to the elementary $n$-tier architecture $\mT_n$.
\end{theorem}

\begin{proof}
Lemma \ref{lem-n-tier} already proves the '$\Rightarrow$' direction. \\
For the '$\Leftarrow$' direction we discriminate three cases and proceed in an indirect manner. \\
\emph{$n = 2$.} Trivial. \\
\emph{$n = 3$.} Let $\mA$ be an architecture which is \emph{not} an $n$-tier architecture and assume, contrary to the theorem, that there exists a homomorphism $h:\mA \mapsto \mT_n$. Then let $C_i = h^-(i)$ be the pullback of $h$. Because $h$ is surjective, non-empty pullbacks $C_i$ exist for $i=1,...,3$. Because $\mA$ is not a 3-tier architecture, there exist elements $c_i^{(j)} \in C_i$ with $h(c_i^{(j)}) = i$ and relations $R_{\alpha} \in \mathbf{R}$ with
\begin{equation*}
	R_{\mu}(c_1^{(1)}, c_2^{(1)}), \ R_{\nu}(c_2^{(2)}), c_3^{(1)}) 
								\text{ and } R_{\sigma}(c_1^{(2}), c_3^{(2)}). 
\end{equation*}
As $h$ is a homomorphism we have 
\begin{equation*}
R_{\sigma}(c_1^{(2)}, c_3^{(2)}) \ \Rightarrow \ 
			h(R_{\sigma})\left( h(c_1^{(2)}), h(c_3^{(2)}) \right) 	\ = T^L(1,3), 
\end{equation*}
which is a contradiction. \\
\emph{$n \ge 4$.} As in the case $n=3$ but note that in addition to the cyclic pattern given above we also might encounter the extended form of this pattern with
\begin{equation*}
	R_{\mu}(c_{i-1}^{(1)}, c_i^{(1)}), \ R_{\nu}(c_i^{(2)}), c_{i+1}^{(1)}) 
						\text{ and } R_{\sigma}(c_i^{(2}), c_k^{(2)}) \text{ with } | i-k | > 1. 
\end{equation*}
As $h$ is a homomorphism we have 
\begin{equation*}
R_{\sigma}(c_i^{(2)}, c_k^{(2)}) \ \Rightarrow \ 
			h(R_{\sigma})\left( h(c_i^{(2)}), h(c_k^{(2)}) \right) 
					\ = T^L(i,k)  \text{ with } | i-k | > 1.
\end{equation*}
which, again, is a contradiction.
\end{proof}

\subsubsection{A ``No Go'' Theorem on Modularity at the Syntax Level}		
							\label{sec-modularity-at-syntax-level}

In section \ref{sec-modules-and-modularity} we have introduced the concept of a generic modular architecture in terms of being homomorphic to an elementary modular architecture $\mM_n$. 
Based on the existance of stand-alone definition of (arbitrary) $n$-tier architectures 
(def.~\ref{def-n-tier-arch}) one might now be induced to also look for a definition of a generic 
modular architecture independent of the existence of a homomorphism to $\mT_n$.

The following theorem, sadly, proves that this search is futile.

\begin{theorem}[Modularity No Go Theorem]		\label{th-generic-modular-architectures}
Every B\&C architecture $\mA = \langle A, \mathbf{R}^A \, \rangle$ is a modular architecture with $n$ modules for arbitrary $n$ provided $n \le |A|$.
\end{theorem}

\begin{proof}
Let $\mA = \langle A, \mathbf{R}^A \, \rangle$ be an arbitrary B\&C architecture and choose $n \le |A|$. Furthermore, let $\{ A_1, A_2,\ldots,A_n \}$ be an (arbitrary!) partition of $A$ into $n$ disjunct sets.
Then write the elementary modular architecture of $n$ elements $\mM_n = \langle M_n, \{ D^M \} \rangle$ and define the relation $D^M \subseteq M_n \times M_n$ as follows:
\begin{equation}		\label{eq-def-DM}
\forall a_i \in A_i \ \forall a_j \in A_j \ \forall \, R^A \in \mathbf{R}^A :
			R^A(a_i, a_j) \Rightarrow D^M(m_i, m_j).
\end{equation}
Then define the homomorphism $h:\mA \to \mM_n$ (with $h = \langle h_0, h_r \rangle$) as
\begin{align}
	h_0 : A				& \to M,											\\
		  a \in A_i 	& \mapsto m_i	\qquad \forall a \in A_i,	\notag 	\\[6pt]
	h_R : \mathbf{R}^A	& \to \{ D^M \},										\\
		  R^A			& \mapsto D^M	\qquad \forall \, R^A \in \mathbf{R}^A.	\notag
\end{align}
Note that we also allow $i = j$.

It now remains to be shown that condition \ref{equ-hom-r} of definition \ref{def-homomorphism} is fulfilled. Take an arbitrary $R^A \in \mathbf{R}^A$ and any $a_i \in A_i$ and $a_j \in A_j$. Then by 
eqn.~\ref{eq-def-DM} we can write:
\begin{align*}
	R(a_i, a_j)	& \Rightarrow D^M(m_i, m_j)		\\
				& = h_R(R^A)\big(m_i, m_j\big)			\\
				& = h_R(R^A)\big(h_0(a_i), h_0(a_j) \big).
\end{align*}
which is all we need.
\end{proof}

We want to stress the consequences of this (null-)result: Theorem \ref{th-generic-modular-architectures} 
shows that we can (\emph{theoretically correct}) ``slice'' any (B\&C) architecture into arbitrary (!) many modules consisting of any (!) collection of elements of the architecture's universe if we do not restrict the (dependency) relation between the resulting modules.
This proves that one cannot sensibly define \emph{modularity} on the syntactic level alone but needs to refer to additional \emph{semantics} in order to be able to do so.
The result, in hindsight, partly justifies the approach in \cite{Efatmaneshnik-2020} who argue that one should define modules in relation to (nonfunctional) requirements the modules fulfill or satisfy. However, the restriction to nonfunctional requirements only cannot be made plausible on the level of syntax as we are equally well and easy able to define modularization with regards to functional requirements (cf.\ our definition \ref{def-functional-modularization}).

Interestingly, this is in stark contrast to the situation of $n$-tier architectures where the syntactical restriction of the relation between the individual tiers is sufficiently strong to allow non-trivial instantiations of the concept.

\section{Discussion}
\label{sec-Discussion}

\subsection{Applicability}

We show that our formal definition of the mathematical syntax of architectures encompasses most (if not all) elements and constructs present in today's architectural standards and architectures produced and consumed by practitioners.

Furthermore, we demonstrate that our approach is an excellent complement to recent foundational work \cite{Dickersen-2020, Wilkinson-2018}, which focuses more on the semantics of architectures than on its syntax. Ref.~\cite{Wilkinson-2018} implicitly uses the syntax of \emph{conceptual structures} \cite{Sowa-1984} on a superficial level while ref.~\cite{Dickersen-2020} uses first order predicate calculus. Our approach is then used to improve and rigorously reconstruct on the syntactical (mathematical) level a theory on modularization as put forward in \cite{Efatmaneshnik-2020}. 

Capitalizing on our notion of architecture \emph{isomorphism}, we then use our formalism to rigorously prove certain relations between architectures. On the formal side, we establish that the naive view that every ``boxes and connectors'' diagram of an architecture (represented rigorously as a \emph{graph}) is fully warranted on the syntactical level. On the categorical side, we prove a classification theorem on general $n$-tier architectures, namely that a boxes and connectors architecture is an $n$-tier architecture if and only if it is homomorphic to an elementary $n$-tier architecture.

However, we are clearly aware that this paper can only capture the very beginning of what we believe could be a completely novel and fruitful way of reasoning about architectures in a mathematically rigorous way.

\subsection{Relation to Model Theory}

The relation of our definition of an \emph{architecture} to the model theoretic concepts of an ($L$-)\emph{model} or an ($L$-)\emph{structure} over a signature $L$ is evident (cf. \cite{Doets-1996, Hodges-1993}). However, we note the following differences:
\begin{itemize}
\item We cannot refer to a common \emph{signature} $L$ shared between (sets of) architectures.
\item Contrary to an ($L$-)structure, we do not have singled out some elements of our (architectural) universe as \emph{constants}.
\item Our definition of (architecture) \emph{homomorphisms} is generalizing the notion used in standard model theory. Because we cannot rely on the fact that a common signature $L$ implicitly "fixes" relations and functions in two different architectures, we need to provide this mapping explicitly through the functions $h_R$ and $h_F$ (cf.~def.~\ref{def-homomorphism}).
\end{itemize}

\subsection{Further Work}

One possible way to extend the work of this paper is to define \emph{refinements} of architectures (like $\mA \succ \mB$) as another (order) relation besides the sub-architecture property ($\mA \subset \mB$). We could, in principle, also differentiate between a "refinement" of architectural elements (artefacts) $a \in A$ and of relations ($R \in \mathbf{R}$) \cite{Broy-1995, Peng-2002}. Thereby we could give a precise meaning to the various forms of ``hierarchicalization'' in architectural work. 

We could also give a precise meaning of the semantics of a given architecture by linking our constructs $\mA$ to $L$-structures and $L$-models over a given signature $L$. The \emph{signature} $L$ could, for instance, then be related to some architectural description languages (ADL).

One can also further develop the properties of the category \textbf{Arch} and transfer results from Category Theory to the architecture domain. Finally, the explicit recognition and inclusion of \emph{names} in our formalism (e.g.~for relations) can be explored in more detail to increase the theory's relevance to the field.

%
\section*{Acknowledgements}
The author gratefully acknowledges the support of Harald Sch\"oning, Software AG's \emph{Vice President Research} and the intensive (mathematical) discussions especially on the very first version of this paper. Additionally, I would like to thank Burkhard Hilchenbach, \emph{Lead Architect Hybrid}
in Software AG's CTO Office, for his comments and pointers to additional topics of high relevance to (true) architects. Finally, I am indebted to my wife, Susi, for repeatedly proof-reading the various versions of this work. Being a biologist, she could only improve the English language, though, so all the conceptual and mathematically mistakes are solely mine \texttt{;-).}

%
\bibliographystyle{splncs04}

\bibliography{Architecture}

\begin{thebibliography}{10}
\providecommand{\url}[1]{\texttt{#1}}
\providecommand{\urlprefix}{URL }
\providecommand{\doi}[1]{https://doi.org/#1}

\bibitem{Agrawal-2009}
Agrawal, A.: Product networks, component modularity and sourcing. Journal of
  Technology Management \& Innovation  \textbf{4}(1),  59--81 (May 2009),
  \url{https://www.jotmi.org/index.php/GT/article/view/art105}

\bibitem{Altoyan-2017}
Altoyan, N., Perry, D.E.: Towards a well-formed software architecture analysis.
  In: Proceedings of the 11th European Conference on Software Architecture:
  Companion Proceedings. p. 173–179. ECSA ’17, Association for Computing
  Machinery, New York, NY, USA (2017). \doi{10.1145/3129790.3129813}

\bibitem{Asha-2020}
Asha, H., Shantharam, N., Annamma, A.: Formalization of {SOA} concepts with
  mathematical foundation. Int. J. Elec. \& Comp. Eng.  \textbf{10}(4),
  3883--3888 (2020). \doi{10.11591/ijece.v10i4.pp3883-3888}

\bibitem{Astesiano-2003}
Astesiano, E., Reggio, G., Cerioli, M.: From formal techniques to well-founded
  software development methods. In: Aichernig, B.K., Maibaum, T. (eds.) Formal
  Methods at the Crossroads. From Panacea to Foundational Support: 10th
  Anniversary Colloquium of UNU/IIST, the International Institute for Software
  Technology of The United Nations University, Lisbon, Portugal, March 18-20,
  2002. Revised Papers, pp. 132--150. Springer, Berlin, Heidelberg (2003).
  \doi{10.1007/978-3-540-40007-3\_9}

\bibitem{Barr-1989}
Barr, M., Wells, C.: Category Theory for Computer Science. Prentice-Hall (1989)

\bibitem{Bass-2013}
Bass, L., Clements, P., Kazmann, R.: {S}oftware {A}rchitecture in {P}ractice.
  SEI Series in Software Engineering, Pearson Education, Upper Saddle River:
  NJ, 3rd edn. (2013)

\bibitem{Broy-1995}
Broy, M.: Mathematical system models as a basis of software engineering. In:
  van Leeuwen~J. (ed.) Computer Science Today, vol.~1000, pp. 292--306.
  Springer, Berlin, Heidelberg (1995). \doi{10.1007/BFb0015250}

\bibitem{Carnap-1934}
{Carnap}, R.: Die Logische Syntax der Sprache. Springer (1934)

\bibitem{Chiang-2020}
{Chiang}, M., {Low}, S.H., {Calderbank}, A.R., {Doyle}, J.C.: Layering as
  optimization decomposition: A mathematical theory of network architectures.
  Proceedings of the IEEE  \textbf{95}(1),  255--312 (2007).
  \doi{10.1109/JPROC.2006.887322}

\bibitem{Chomsky-1975}
Chomsky, N.: The Logical Syntax of Linguistic Theory. Berlin, Springer (1975)

\bibitem{deBoer-2004}
{de Boer}, F.S., {Bonsangue}, M.M., {Jacob}, J., {Stam}, A., {van der Torre},
  L.: A logical viewpoint on architectures. In: Proceedings. Eighth IEEE
  International Enterprise Distributed Object Computing Conference, 2004. EDOC
  2004. pp. 73--83 (2004)

\bibitem{Dickersen-2020}
{Dickerson}, C.E., {Wilkinson}, M., {Hunsicker}, E., {Ji}, S., {Li}, M.,
  {Bernard}, Y., {Bleakley}, G., {Denno}, P.: Architecture definition in
  complex system design using model theory. IEEE Systems Journal pp. 1--14
  (2020). \doi{10.1109/JSYST.2020.2975073}

\bibitem{Doets-1996}
Doets, K.: Basic model theory. Studies in logic, language and information, CSLI
  Publications, Stanford, MA (1996)

\bibitem{Dori-2019}
Dori, D., Sillitto, H., Griego, R.M., McKinney, D., Arnold, E.P., Godfrey, P.,
  Martin, J., Jackson, S., Krob, D.: System definition, system worldviews, and
  systemness characteristics. IEEE Systems Journal pp. 1--11 (2019).
  \doi{10.1109/JSYST.2019.2904116}

\bibitem{Efatmaneshnik-2020}
{Efatmaneshnik}, M., {Shoval}, S., {Qiao}, L.: A standard description of the
  terms module and modularity for systems engineering. IEEE Transactions on
  Engineering Management  \textbf{67}(2),  365--375 (2020).
  \doi{10.1109/TEM.2018.2878589}

\bibitem{Fiadeiro-1996}
Fiadeiro, J.L., Maibaum, T.: A mathematical toolbox for the software architect.
  In: Proceedings of the 8th International Workshop on Software Specification
  and Design. p.~46. IWSSD ’96, IEEE Computer Society, USA (1996).
  \doi{10.1109/IWSSD.1996.501146}

\bibitem{Garlan-2003}
Garland, D.: Formal modeling and analysis of software architecture: Components,
  connectors, and events. In: Bernardo, M., Inverardi, P. (eds.) Formal Methods
  for Software Architectures: Third International School on Formal Methods for
  the Design of Computer, Communication and Software Systems: Software
  Architectures, SFM 2003, Bertinoro, Italy, September 22-27, 2003. Advanced
  Lectures, pp. 1--24. Springer, Berlin, Heidelberg (2003).
  \doi{10.1007/978-3-540-39800-4\_1}

\bibitem{Goguen-1989}
Goguen, J.A.: A categorical manifesto. Technical Manifesto PRG-72, Oxford
  University Computing Laborartory, Programming Research Group (March 1989)

\bibitem{Goikoetxea-2004}
Goikoetxea, A.: A mathematical framework for enterprise architecture
  representation and design. Int. J. Inf. Technol. Decis. Mak.  \textbf{3},
  5--32 (2004)

\bibitem{Hamburger-1987}
Hamburger, K.: Die Logik der Dichtung. Klett, Stuttgart (1987)

\bibitem{SAG-2020}
{Hilchenbach}, B., {Strnadl}, C.F.: Software {AG} reference architecture -
  capabilities view.
  https://www.softwareag.cloud/site/reference-architecture/software-ag-reference-architecture.html
  (2020), last accessed April, 6th, 2020

\bibitem{Hodges-1993}
Hodges, W.: Model Theory. Encyclopedia of Mathematics and its Applications,
  Cambridge University Press, Cambride, MA (1993)

\bibitem{ISO-42010}
ISO: {ISO/IEC/IEEE} 42010:2011: Systems and software engineering —
  architecture description. Standard, International Standards Organization,
  Geneva, CH (2011)

\bibitem{Lankhorst-2017}
{Lankhorst et al.}, M.: Enterprise Architecture at Work. Springer, Berlin,
  Heidelberg (2017). \doi{10.1007/978-3-662-53933-0}

\bibitem{Leinster-2014}
Leinster, T.: Basic Category Theory., Cambridge Studies in AdvancedMathematics,
  vol.~143. Cambridge University Press, Cambridge, UK (2014)

\bibitem{Lopez-2009}
L\'{o}pez-Sanz, M., Vara, J.M., Marcos, E., Cuesta, C.E.: A model-driven
  approach to weave architectural styles into service-oriented architectures.
  In: Proceedings of the First International Workshop on Model Driven Service
  Engineering and Data Quality and Security. p. 53–60. MoSE+DQS ’09,
  Association for Computing Machinery, New York, NY, USA (2009).
  \doi{10.1145/1651415.1651426}

\bibitem{Mitchell-1990}
Mitchell, W.J.: The Logic of Architecture. The MIT Presse, Cambridge, MA (1990)

\bibitem{OCL}
{OMG}: {OMG Object Constraint Language}. Standard, The Object Management Group
  (2014), \url{http://www.omg.org/spec/OCL/2.4}

\bibitem{Peng-2002}
Peng, J., Abdi, S., Gajski, D.: Automatic model refinement for fast
  architecture exploration [soc design]. In: Proceedings of {ASP}-{DAC}/{VLSI}
  Design 2002. 7th Asia and South Pacific Design Automation Conference and 15h
  International Conference on {VLSI} Design. pp. 332--337. IEEE (2002).
  \doi{10.1109/ASPDAC.2002.994944}

\bibitem{Sowa-1984}
Sowa, J.F.: Coneptual Structures: Information Processing in Mind and Machine.
  Addison-Wesley, Reading, MA (1984)

\bibitem{Strnadl-2006}
{Strnadl}, C.F.: Aligning business and {IT}: The process-driven architecture
  model. IS Management  \textbf{23}(4),  67--77 (2006)

\bibitem{UML}
{The Object Management Group}: {OMG Unified Modeling Language}. Standard, The
  Object Management Group (2017), \url{https/www.omg.org/spec/UML/}

\bibitem{SysML}
{The Object Management Group}: {OMG} {S}ystems {M}odeling {L}anguage.
  Standard~1.6, The Object Management Group (2019),
  \url{https://www.omg.org/spec/SysML/1.6/}

\bibitem{TOGAF}
{The Open Group}: {The TOGAF Standard Version 9.2}. Standard, The Open Group
  (2018)

\bibitem{Archimate}
{The Open Group}: {Archimate 3.1 Specification}. Standard~3.1, The Open Group
  (2019), \url{https://pubs.opengroup.org/architecture/archimate3-doc/}

\bibitem{Wermelinger-1999}
Wermelinger, M., Fiadeiro, J.L.: Algebraic software architecture
  reconfiguration. In: Proceedings of the 7th European Software Engineering
  Conference Held Jointly with the 7th ACM SIGSOFT International Symposium on
  Foundations of Software Engineering. p. 393–409. ESEC/FSE-7,
  Springer-Verlag, Berlin, Heidelberg (1999)

\bibitem{Whitehead-Russell-1910}
{Whitehead}, A.N., {Russell}, B.: Principia Mathematica. Cambridge University
  Press, Cambridge, UK (1910-1913)

\bibitem{Wilkinson-2018}
Wilkinson, M.K., Dickerson, C.E., Ji, S.: Concepts of architecture, structure
  and system (2018), \url{https://arxiv.org/abs/1810.12265}

\bibitem{Wittgenstein-21}
{Wittgenstein}, L.: Logisch-philosophische {A}bhandlung. Wilhelm Ostwalds
  Annalen der Naturphilosphie  \textbf{4}(3-4) (1921)

\bibitem{Yanchuk-2006}
Yanchuk, A., Alexander, I., Maurizio, M.: Towards a mathematical foundation for
  service-oriented applications design. J. Software  \textbf{1}(1),  32--39
  (2006). \doi{10.4304/jsw.1.1.32-39}

\end{thebibliography}

\end{document}